\documentclass[usenatbib,usegraphicx]{mn2e}
\begin{document}

\title[Self-interacting dark matter haloes]{Formation and Evolution of Self-Interacting Dark Matter Haloes}

\author[K. Ahn \& P. R. Shapiro]{Kyungjin Ahn\thanks{Email:
    kjahn@astro.as.utexas.edu} and Paul R. Shapiro\thanks{Email:
    shapiro@astro.as.utexas.edu} \\ 
Department of Astronomy, The University of Texas at Austin, 1
University Station C1400, Austin, TX 78712, USA}

\maketitle
\begin{abstract}
We have derived the first,
fully-cosmological, similarity solutions for cold dark matter (CDM)
halo formation in the
presence of nongravitational collisionality (i.e. elastic scattering), 
which provides an analytical theory of the 
effect of the self-interacting dark matter (SIDM)
hypothesis on halo density profiles. Collisions 
transport heat inward which flattens the central cusp of the CDM
density profile to produce a constant-density core,
while continuous infall pumps energy into the halo to
stabilize the core against gravothermal catastrophe. This is contrary to
previous analyses based upon isolated haloes, which predict core
collapse within a Hubble time. These solutions
improve upon earlier attempts to model the formation and evolution of
SIDM haloes, offer deeper insight than existing N-body experiments, and
yield a more precise determination of the dependence of halo density
profile on the value of the CDM self-interaction cross section. 
Different solutions arise for different values of the dimensionless
collisionality parameter $Q\equiv \sigma \rho_{b} r_{\rm vir} \propto
r_{\rm vir}/\lambda_{\rm mfp}$, where $\sigma$ is the scattering cross
section per unit mass, $\rho_{b}$ is the cosmic mean matter density, 
$r_{\rm vir}$ is halo virial radius and $\lambda_{\rm mfp}$ is the
collision mean free path. The maximum flattening of central density
occurs for an intermediate value of $Q$, $Q_{\rm th}$, at which the
halo is maximally relaxed to isothermality.
The density profiles with constant-density cores preferred by dwarf
and low surface brightness galaxy (LSB)
rotation curves are best fit by the maximally-flattened ($Q=Q_{\rm th}$)
solution. If we assume that dwarfs and LSB galaxies formed at their
typical collapse epoch in $\Lambda$CDM, then the value of $\sigma$
which makes $Q=Q_{\rm th}$ is $\sigma \simeq 200 \,\rm{cm^{2}
\,  g^{-1}}$, much higher than previous estimates, $\sigma \simeq
[0.5-5] \,\rm{cm^{2} \, g^{-1}}$, based on N-body experiments.
If $\sigma$ is independent of collision velocity, then the same
value $\sigma \simeq 200 \,\rm{cm^{2}\,g^{-1}}$ would make
$Q>Q_{\rm th}$ for clusters, which typically formed only recently,
resulting in relatively less flattening of their central density profile and a
smaller core.
\end{abstract}

\section{Introduction}
\label{intro_sect}
The cold dark matter (CDM) model provides a successful framework for
understanding the formation and evolution of structure in our
universe.
According to this model, gravity amplifies primordial density
fluctuations, and structure forms by hierarchical clustering: small
objects form first, later merging to form larger objects. The most
promising candidate for CDM is the weakly interacting massive particle
(WIMP). In this picture, the microscopic interaction between CDM particles
is negligible (collisionless), and they interact only by
gravity. However, this assumption has not been fully verified and it
is fair to say that the microscopic nature of CDM is still unknown.
It is important, therefore, to explore the consequences of varying this
underlying assumption about CDM in the hope that astronomical
observations can be used to place meaningful constraints. 

The possible variation of the microscopic nature of cold dark matter
is closely linked to the problems of the CDM model. Despite its
success, the CDM model has several problems which exist mostly in the
small scale regime (see, for example,
\citealt{moorereview}). Among these problems, much attention has been
focused on the N-body simulation results for the inner density slope,
since the observed rotation curves of dark-matter dominated dwarf and
low surface brightness (LSB) disk galaxies tend to favour mass profiles
with a flat-density core unlike the singular profiles of the CDM
N-body simulations (e.g. \citealt{fp}; \citealt{march}).
The latter are generally characterized by an empirical fitting formula
for the spherically averaged density profiles in those N-body results,
either the 
Navarro-Frenk-White (NFW) profile (\citealt{nfw}), for which $\rho\propto
r^{-1}$ as $r\rightarrow 0$, or the Moore profile
(\citealt{moore}), for which $\rho\propto r^{-1.5}$ 
instead\footnote{Recently, \citet{hayashi} and \citet{netal}
reported that the logarithmic slope of the spherically averaged
density profile of dark matter haloes in their $\Lambda$CDM
simulations, $\ -{\rm d}\ln \rho / {\rm d} \ln r $,
decreases monotonically towards the centre. This change of slope,
however, does not seem to fully account for the observed rotation curves. 
\citet{dms} conclude, by surveying several simulation results by different
groups, that neither the typical NFW profile nor the Moore profile
provides a good fit, but one needs a little more complex form such as
the one by \citet{netal}. }.
It was controversial whether the observed data was resolved
well enough to indicate a soft core (e.g. see \citealt{vs} for possible 
beam smearing effect on the data), but observations have built up which
favour the soft core even after eliminating the beam smearing effect
(e.g. \citealt{march} and references therein). After much more work on
the N-body simulations of CDM, the discrepancy between these data and
the numerical halo profiles remains significant (see, for instance,
\citealt{dms}) 

This apparent discrepancy between the N-body simulation results and
observations of dwarf galaxy rotation curves
has raised a question about the nature of dark matter, including the
assumption that CDM is 
collisionless. People have suggested
solutions to this discrepancy which either preserves the collisionless
nature of the dark matter or else adopts a new picture. In the former
category are explanations which 
attribute the discrepancy to astrophysical processes beyond the pure
N-body dynamics
(e.g. \citealt{esh,wk}) or to a primordial power spectrum tilted away from the
Harrison-Zel'dovich shape (e.g. \citealt{zb}).  
In the latter category, the proposal of self-interacting dark matter (SIDM) by
\citet{ss} has received a lot of attention. 
In this picture, microscopic
interaction between dark matter particles is non-negligible and can
affect the dynamics of halo formation. 
Since the actual identity and microscopic nature of CDM is
still unknown, it is important to explore the consequences of 
such hypotheses in the hope that astronomical observations can be used
to place meaningful constraints.

The truncated isothermal sphere (TIS) halo model developed by 
\citet[][also see \citealt{tis2}]{tis1} provides a simple physical clue about
the existence of such soft cores in haloes of cosmological origin which
otherwise closely resemble the CDM haloes found by N-body simulation in all
respects except the presence of the inner cusp.
The model is based upon the assumption 
that haloes form from the collapse and virialization of 
``top-hat'' density perturbations and are spherical, isotropic, and
isothermal. This leads to a unique, nonsingular TIS, a minimum-energy
solution of the Lane-Emden equation. 
%The size $r_{t}$ and velocity dispersion $\sigma_{V}$ are unique
%functions of the mass and redshift of formation of the halo for a
%given background universe. The TIS density profile flattens to a
%constant central value, $\rho_{0}$, which is roughly proportional to
%the critical density of the universe at the epoch of collapse, with a
%small core radius $r_{0}\simeq r_{t}/30$ (where 
%$\sigma_{V}^2=4\pi G \rho_{0} r_{0}^2$ and $r_{0}\equiv r_{\rm
%{King}}/3 $, defined by \citealt{tis1}, p. 228). 
The resulting density
profile is nearly indistinguishable from that deduced from the
observed rotation curves of dwarf and LSB disk galaxies, as fit by
\citet{bur1} with a profile involving a soft core
(\citealt{tis2}). 
%The same TIS mass 
%profile agrees well with the fit to CDM N-body simulations by NFW
%(i.e. fractional deviation of $\sim 20\%$ or less) at all radii
%outside of a few TIS core radii (i.e. outside the King radius or so),
%for NFW concentration parameters $4\le c_{\rm{NFW}} \le 7$
%(\citealt{tis2}). The flat density core of the TIS 
%halo differs from the singular cusp of the NFW profile at small radii,
%but this involves only a small fraction of the halo mass, thus not
%affecting their good agreement outside the core. 
The assumption of
isothermality which underlies the TIS halo model is the primary reason
that the central cusp of the CDM N-body haloes is replaced by a soft
core. The N-body haloes are also approximately isothermal, but there is
typically a small temperature dip near the centre. This suggests that
if some mechanism existed to transport heat inward so as to make CDM
haloes more isothermal, they might exhibit soft cores while maintaining
the same basic halo structure at larger radii already found for CDM
haloes. 
The collisionality of SIDM serves as such a mechanism: elastic collisions
transport heat inward, which flattens the central cusp of the CDM
density profile to produce a constant-density core.

The problem of SIDM halo formation has so far been studied primarily
by numerical N-body experiments. Some of the first attempts to
calculate the effect of the elastic scattering of SIDM on halo
structure formation involved {\it isolated} haloes which were assumed
initially to 
follow the equilibrium profiles (e.g. NFW profile) found in
collisionless N-body simulations of standard CDM (\citealt{bur2};
\citealt{kw}; both start from the Hernquist profile
(\citealt{hernquist}) which is a good
approximation to the NFW profile). \citet{kw}, for instance, found
that SIDM haloes can form flat density cores within a relaxation time
as expected, but also that the lifetime of such flat cores is only a
few relaxation times. They concluded, therefore, that most galactic
haloes would have undergone core collapse.

The effect
of SIDM collisionality on halo structure has also
been studied by \citet[BSI~hereafter]{bsi}
by solving 1D, quasi-static fluid equations.
These authors also considered {\it isolated} haloes like those in the N-body
experiments mentioned above,
adopting 1D, spherical symmetry, with non-cosmological
boundary conditions. They treated the dynamics of SIDM by a fluid
approximation developed previously in the study of stellar dynamics,
derived from the Boltzmann equation, in which they modified the heat
conduction term to handle the elastic scattering of the SIDM
particle-particle interaction. They solved the spherically-symmetric,
virialized ``gravothermal fluid'' equations of \citet{le}, which
include mass conservation, hydrostatic equilibrium, an equation for
heat conduction, and the first law of thermodynamics. According to
these equations, the halo is time-dependent because heat conduction
causes it to evolve through a continuous sequence of hydrostatic
equilibria. An analytical self-similar solution to these equations was
found in the limit of large scattering mean free path -- the limit
where the mean free path is much larger than the size of the halo -- following
the derivation of \citet{le} for globular clusters. The BSI solution
shows that secular evolution always takes a configuration in the long
mean free path limit and drives it into the short mean free path
regime. In order to track the evolution 
into the short mean free path regime, as well, BSI then used their similarity
solution as the initial condition for a numerical solution of the same
gravothermal fluid equations, but for finite scattering cross-section.
This approach made it possible to
follow core collapse to a much more advanced stage than the N-body
experiments could. From this, they concluded that the ultimate core
collapse time was much larger than the relaxation time in the core,
long enough even to exceed a Hubble time in some cases\footnote{In a
follow-up paper, the authors showed how this process could lead 
to the formation of seeds for super-massive black
holes (\citealt{bs}).}.
Their
estimated core collapse time is $t_{\rm coll}\simeq 290 \, t_{r}$, which
contradicts the result found by \cite{bur2} and
\cite{kw} that the core collapse time was only a few relaxation times.

This apparent discrepancy in core collapse timescale between the
study by BSI and the numerical N-body experiments (\citealt{bur2};
\citealt{kw}) may be attributed to the fact that 
their adopted initial conditions
were different. In BSI, as mentioned, the initial condition was tuned
to be in the extremely long mean free path limit, $\lambda_{\rm
  mfp}/H \gg 1$, 
where $\lambda_{\rm mfp}$ is the mean free path and $H$ is the
gravitational 
scale height or roughly the halo size. This occurs when the
system is either dilute enough or the scattering cross section (per
unit mass)
$\sigma$ is small enough, since 
$\lambda_{\rm mfp} \propto 1/(\sigma \rho)$.
According to BSI, most of the collapse time
is spent to reach the condition $\lambda_{\rm mfp}/H\simeq 1$, and
the halo density profile always has a flat core.
By contrast, \citet{kw} started with a cuspy profile with
parameters which corresponded to
a condition $\lambda_{\rm mfp}/H \simeq 0.1 - 3.0$. In this case, SIDM halo
cores can quickly reach $\lambda_{\rm mfp}/H \simeq 1$ or they are already in
the short mean free path limit, which then requires only a few relaxation
times for core collapse.

In what follows, we will cover the whole range of $\sigma$ and 
show that the observed dwarf-galaxy rotation curves are best-fit when
the SIDM interaction has its
maximal effect, which occurs when $\lambda_{\rm mfp}/H \simeq 1$,
so the regime of greatest interest may be that of
\citet{kw}. If so, then most isolated haloes would, indeed, suffer core collapse
within a Hubble time. However, the shared limitation of the analyses of
BSI, \citet{bur2} and \citet{kw}, that of non-cosmological boundary
conditions, is a severe one. Cosmological infall may inhibit core collapse.
If cosmological infall can delay core collapse substantially,
previous estimates based on isolated halo models would change by shifting the
time of the onset of core collapse until cosmological infall becomes
negligible. 

The effect of cosmological boundary conditions on
SIDM halo formation has been studied numerically by
cosmological N-body simulations, in which Gaussian random noise
initial conditions for CDM were incorporated. Early work along these
lines attempted to derive the maximal effect of
collisionality by adopting the fully collisional limit which corresponds
to ordinary gas dynamics (\citealt{yoshida-strong}; \citealt{moore-sidm}).
The surprising result they reported was that simulations yielded
density profiles with central cusps even steeper -- with logarithmic
slope close to $-2$ -- than those in collisionless
N-body simulations. Subsequent cosmological N-body experiments which
treated the SIDM elastic scattering in more detail, however, reported
that halo density profiles were flatter than those of either fully
collisional or purely collisionless simulations
(\citealt{yoshida-weak}; \citealt{dave}; \citealt{cavf}).
They found that values of $\sigma$ in the range $\sigma \simeq [0.1
- 5] \,{\rm cm}^{2} {\rm g}^{-1} $ (the range of preferred
$\sigma$ varies among
these works, but within less than an order of magnitude)
produced SIDM haloes with sufficient profile flattening to account for
dwarf galaxy rotation curves,
which did not suffer from the rapid core collapse
identified in the earlier N-body experiments for isolated
haloes. \citet{dave} speculated that this might be the result of
cosmological infall, which was absent from the calculations of
isolated haloes. In addition, \citet{yoshida-weak} found that 
SIDM collisions in cluster-sized
haloes would be more frequent than in dwarf-sized haloes,
thus producing relatively larger cores in clusters, 
while observations tend to show that
dwarfs and LSBs show relatively larger cores than clusters. This
led them to suggest that the scattering cross section be
velocity-dependent ($\sigma\propto 1/v$) so that more
massive haloes would have a smaller degree of density profile
flattening. Later, \citet{cavf} tested
this hypothesis and confirmed that it could match observed core sizes
from dwarfs to galaxy clusters.

Further study of the formation and evolution of SIDM haloes is
warranted to resolve the issues raised by previous work and put the
subject on a firmer theoretical footing. On the one hand, the
numerical N-body simulations and 1D semi-analytical treatment mentioned
above of isolated haloes with non-cosmological boundary conditions are
unable to address the important effects of cosmological infall. On the
other hand, the fully cosmological N-body simulations which have been
performed of haloes that arise during large-scale structure formation
in the SIDM model have so far been limited by numerical
resolution and dynamic range. As a result, simulation results
published to date do not attempt a detailed enough comparison with
observed galactic rotation curves to determine if SIDM haloes can
really match them or to give a very reliable constraint on the SIDM
cross section. Such simulations do not afford enough insight into the
underlying dynamical processes which govern the halo structure,
either. Finally, the wide range of typical collapse epochs expected in
a CDM universe for the wide range of halo masses extending from dwarf
galaxies to clusters suggests that the effect of the SIDM interaction
may be halo-mass-dependent; this dependence has not yet been
adequately explored.

Towards this end, we have derived a fully cosmological model for
the origin and evolution of CDM haloes in the presence of
nongravitational collisionality (i.e. elastic scattering). We have
combined the fluid approximation of BSI with the spherical infall
model for cosmological perturbation growth to yield fully
time-dependent, detailed similarity solutions for SIDM haloes, for
arbitrary degree of collisionality. We shall apply these solutions to
test the hypothesis that cosmological infall retards the core collapse
of SIDM haloes, and compare the predicted SIDM halo profiles with the
mass profiles inferred from dwarf galaxy rotation curves. This will
enable us to place much better quantitative constraints on the SIDM
cross section and better assess the validity of the SIDM
model. 

Subsequent to the original suggestion by \citet{ss} and the
exploration described above of the halo structure which results,
related work has focused on constraining the SIDM hypothesis 
by its implications for other astrophysical phenomena or 
attributing density flattening to more complicated CDM
dynamics or gas dynamical
feedback effects within the standard CDM picture. There seem to be strong
observational constraints on the possible range of $\sigma$.
\citet{go} ruled out a range of $ \sigma $,
$ \sigma = [0.3 - 10^4]\,{\rm cm}^{2} {\rm g}^{-1} $, based on
their calculation of the evaporation time 
of the dark matter haloes of elliptical galaxies in the clusters.
\citet{natarajan} rule out all values of
$\sigma > 42 \, {\rm cm}^{2} {\rm g}^{-1}$ by comparing the predicted
truncation radii of SIDM haloes inside
clusters by ram-pressure stripping to those of observed haloes 
which they obtain using cluster
gravitational lensing observations. 
\citet{ho} conclude that if $\sigma \gg 0.02 \, {\rm cm}^{2}
{\rm g}^{-1}$,the supermassive black holes in the centres of galactic
haloes would be more massive than observed.
Along the line of explaining flattened cores within
the standard CDM picture 
(see \citealt{primack} for recent review),
\citet{ehpcs}, as in \citet{esh},
investigate the effect of dynamical friction by clumpy substructure
associated with baryonic dissipation
and show that a flat core can be generated. \citet{wk} and
\citet{hwk} claim that soft cores can be induced by the presence of a
CDM bar structure.
\citet{hayashi} argue that the gas rotation speed in a galactic
disk may be different from the dark matter circular velocity, which
would, therefore, be
wrongfully interpreted as implying the existence of soft cores.  
Dynamical feedback effects from supernova explosions is another
possibility (\citealt{nef}).
Observationally, related to the missing satellite problem, the
gravitational lensing flux anomaly seems to require clumpy, dark substructures
(\citealt{mz}; \citealt{dk}; \citealt{kgp}; \citealt{mjow}), even
though no firm 
conclusion has been established yet.

Nevertheless, we believe that SIDM is still a viable candidate for dark
matter. As we shall discuss below in \S \ref{discussion_sect},
the previous analyses restricting $\sigma$ are subject to
significant caveats.
It is meaningful to study this subject because it can shed 
light on the nature of dark matter whose origin we do not know yet,
and simply because CDM problems are far from reaching a firm
conclusion. 
Also, the research effort on SIDM has not yet reached the level of
that on
collisionless CDM. 

We have previously reported the results of this paper in a summarized
form elsewhere. \citet{as1,as2} briefly described the result from \S
\ref{nbody_sidm_sect}. \citet{ourreview} summarized the results in
light of the current status of CDM research. However, we warn the
reader that these works should be considered as guidelines to the
detailed, complete picture of this paper.

The plan of this paper is as follows. 
In \S \ref{ssgc_sect}, we review the previous work on self-similar
gravitational collapse in the spherical infall model.
In \S \ref{fluidapprox_sect}, we show how the 
fluid approximation derived from the Boltzmann equation can be
generalized to include a conduction term which accounts for
nongravitational collisionality in an otherwise collisionless dark
matter gas. The assumptions which underlie this fluid approximation
will be justified by comparison with the results of cosmological
N-body simulations of CDM haloes and with detailed similarity solutions
for halo formation by gravitational collapse of collisionless dark matter.
In \S \ref{ss_sidm_sect}, we show that gravitational collapse by
spherical infall can be self-similar even in the presence of
collisionality, and that the specific condition required for
self-similarity is relevant to galactic haloes. We then describe the basic
equations for the dynamics of SIDM haloes. 
In \S \ref{result_sect}, we present the similarity solution and
its application to test the SIDM hypothesis, including a comparison
with previous N-body results and observed galactic rotation curves. 
Our conclusions and discussion are in \S \ref{discussion_sect}.

\section{SELF-SIMILAR GRAVITATIONAL COLLAPSE}
\label{ssgc_sect}
\subsection{Previous analytical models for halo formation: the
spherical infall model}
\label{sshistory_sect}
Analytical approximations have been developed to model the formation
of haloes by the 1D growth of spherical cosmological density
perturbations, in either a collisionless gas or a fluid. We shall need
to refer to some of these solutions to justify our fluid approximation
in \S \ref{fluidapprox_sect}. We shall also build upon these
earlier solutions in deriving new ones which add the effects of
collisionality in \S \ref{fluidapprox_sidm_sect} and
\ref{ss_sidm_sect}. It is necessary for us to 
begin, therefore, by briefly recounting this earlier work.

\citet{gg} first presented the concept of the so called ``secondary
infall model (SIM)''. 
This SIM refers to the effect of the addition of a point mass to a uniform,
expanding Friedmann-Robertson-Walker universe as a perturbation which
causes the surrounding spherical shells to decelerate relative to the
background universe, until they reach a radius of maximum expansion
and re-collapse. Subsequent work generalized this approach to include
spherically-symmetric initial perturbations for which the overdensity
profile depends upon radius or mass as a scale-free power-law.
Along these lines, \citet[FG hereafter]{fg} studied the dynamics
of collisionless CDM haloes using a self-similar model, adopting a
scale-free initial overdensity parametrized by its shape:
$\varepsilon$ in equation (\ref{overmass}). 
\citet{bert} studied a special case of FG
but also extended the analysis to a collisional
fluid. \citet[HS hereafter]{hs} showed that a power-law power
spectrum would indeed generate a scale-free initial condition,
such as was adopted by FG. 
They then argued that the resulting nonlinear structure
would be described by a power-law profile determined by the shape of
the power spectrum.

Previously mentioned works adopted a rather unrealistic condition 
for the collisionless case, that of purely radial motion.
N-body simulations of CDM halo formation find that the virialized
region tends toward isotropic random velocities.
Some attempts to incorporate tangential velocities within the framework
of spherical symmetry have also been made by \citet{rg}, \citet{afh} and \citet{hbf}.
Along these lines, one may also refer to work by
\citet{kull}, \citet{lh}, \citet{popolo}, \citet{hiotelis} and references therein.

The fluid approximation of collisionless CDM halo formation has emerged
recently. \citet{tca} showed that one could use fluid-like
conservation equations to mimic the radial-only SIMs such as the
FG model. \citet{sco} extended this analysis to include
tangential motion. In these pictures, one could expect, for instance,
accretion shock structure. \citet{tca} and \citet{sco} therefore used a
``shock jump condition'' which one would only expect from a purely
collisional fluid. The fluid approximation has been used in the
literature of stellar dynamics, but its application to CDM halo
dynamics is rather new, and as will be described in \S 
\ref{fluidapprox_sect}, it simplifies the description of 
dark matter dynamics substantially. 

Self-similar spherical infall models have also been used to study the
effect of incorporating additional baryonic physics.
\citet{bert-cool}, \citet{owv}, and \cite{abn} have studied the effect of gas
cooling on galaxy formation using self-similar models. Because the
system was tuned to maintain self-similarity in the presence of
cooling, the cooling function used is not perfectly physical. Nevertheless,
as shown by \citet{abn}, one can use these models at least to test
hydrodynamic codes in the presence of cooling. Moreover, these models
capture the generic behaviour of galactic dynamics in the presence of
realistic cooling.

The similarity model we shall develop in this paper is the first
self-similar halo model which includes the effective heat conduction
resulting from SIDM collisionality. In section \S \ref{fg_sect} and
\ref{hs_sect}, we describe how SIMs with scale-free linear
perturbations arise naturally from the Gaussian random noise
primordial density fluctuations, thus leading to the self-similar
evolution of nonlinear structures.

\subsection{Halo formation from scale-free linear perturbations}
\label{fg_sect}
In the Einstein-de Sitter (EdS) background universe, an initial linear
perturbation whose mass profile is spherically symmetric and has a 
scale-free, power-law form
\begin{equation}
\label{overmass}
\frac{\delta M}{M}\propto M^{-\varepsilon }
\end{equation}
results in structure formation which is self-similar (FG). Each
spherical mass shell around the centre expands until it reaches a
maximum radius (turnaround radius $r_{\rm ta}$), and re-collapses. 
For a given $ \varepsilon  $, we have 
\begin{equation}
\label{rta}
r_{\rm ta}\propto t^{\xi },
\end{equation}
where
\begin{equation}
\label{xi}
\xi =\frac{2}{3}\left( \frac{3\varepsilon +1}{3\varepsilon }\right).
\end{equation}
Since there are no characteristic length or time scales for this
problem other than the turn-around radius $r_{\rm ta}$ and the Hubble
time $t$, the gravitational collapse which ensues from this
scale-free initial condition must be self-similar as long as the
background universe is Einstein-de Sitter, in the absence of physical
processes which introduce additional scales (e.g. SIDM collisionality)

In general, if the
unperturbed matter is a cold fluid, the infall which results from this
perturbation is highly supersonic and is terminated by a strong
accretion shock which thermalizes the kinetic energy of collapse. The
accretion shock radius is guaranteed by self-similarity to be a fixed
fraction of $r_{\rm ta}(t)$ at all times. The mean density of the
postshock region is, therefore, always a fixed multiple of the cosmic
mean matter density. For most cases of interest, this postshock region
is close to hydrostatic. For a collisionless gas, a similar
description applies as long as the infalling matter initially had
small (or zero) random motions. In that case, each mass shell collapses
supersonically as a single stream until it encounters a region of
shell-crossing and density caustics, which encompasses all previously
collapsed (i.e. interior) mass shells. All collapsed mass shells
inside this region oscillate about the centre. The radius of this
region of shell-crossing, given by the outermost density caustic, is
analogous to the shock radius in the fluid case.

\begin{figure}
\includegraphics[width=84mm]{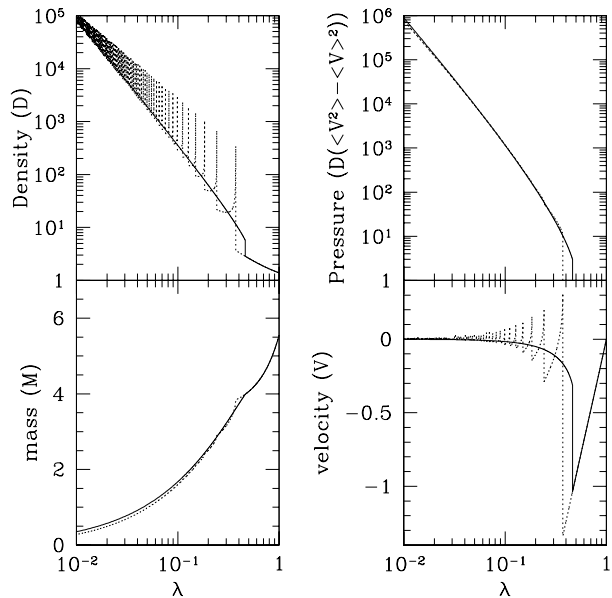}
\caption{Self-similar collisionless halo formation for
$\varepsilon=1$: Comparison of the skewless-fluid approximation to
the exact collisionless Bertschinger solution. Solid lines represent the
solution obtained from the fluid approximation in the radial
direction, while dotted lines represent the collisionless Bertschinger
solution. Spikes in the density plot
simply represent infinite values, corresponding to caustics,
and therefore there is no physical
significance in the height of these spikes. However, spikes in the
velocity plot are finite and real. Note that solid lines do not
represent the $\gamma=5/3$ fluid Bertschinger solution.}
\label{fig-bert}
\end{figure}

\begin{figure}
\includegraphics[width=84mm]{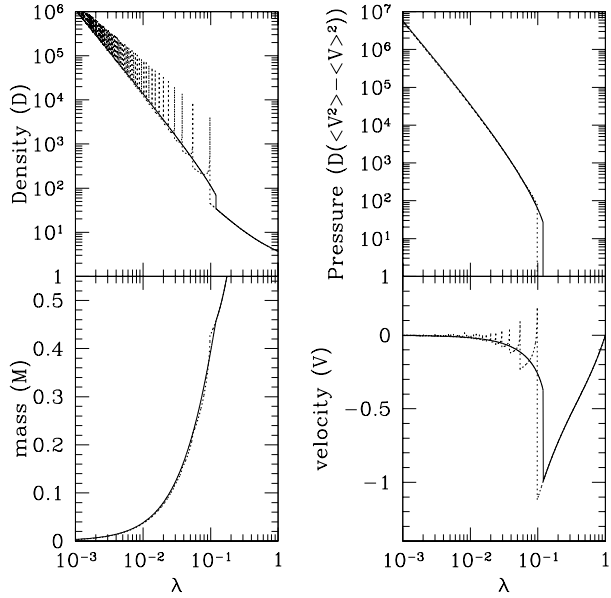}
\caption{Same as Figure \ref{fig-bert}, but $\varepsilon=1/6$.
Note again that the solid line was not generated from the
$\gamma=5/3$ fluid approximation, but rather from the radial-only fluid
approximation described by equations (\ref{mass-bert}) - (\ref{energyr-bert}).
}
\label{fig-e16}
\end{figure}

Results for the purely collisionless case were presented for arbitrary values
of $\varepsilon$ by FG, and for $\varepsilon=1$ by \citet{bert}
(where the latter included a fluid component, as well).
Figures \ref{fig-bert} and \ref{fig-e16} show the exact similarity solutions
for the purely collisionless cases with $\varepsilon=1$ and
$\varepsilon=1/6$, respectively. As we describe below in \S
\ref{hs_sect}, 
these values roughly bracket the range relevant to cosmological haloes
in a CDM universe. We will refer to these solutions again for
comparison in deriving our fluid approximation in \S
\ref{fluidapprox_sect}. 

\subsection{Halo formation from peaks of the Gaussian random noise
primordial density fluctuations}
\label{hs_sect}
The theory of halo
formation from peaks in the density field which result from
Gaussian-random-noise 
initial density fluctuations draws an interesting connection between the
average density profile around these peaks and the shape of the
fluctuation power spectrum.
According to HS, local maxima of Gaussian
random fluctuations in the density can serve as the progenitors of
cosmological structures. They show that rare density peaks ($ \nu
\geq 3 $, where $\nu$ corresponds to $\nu \sigma_{M}$ peak) have a
simple power-law 
profile\footnote{\citet{bbks} also get a similar result: local density
maxima have a triaxial profile,
but as $ \nu  $ increases it becomes more and more spherical with
a profile converging to equation (\ref{overden}).}
\begin{equation}
\label{overden}
\Delta_{0}(r)\propto r^{-(n+3) },
\end{equation}
where $\Delta_{0}(r)$ is the accumulated overdensity
inside radius r, and $n$ is the effective index of the power
spectrum $P(k)$ approximated as a power-law $P(k)\propto k^{n}$ at
wavenumber $k$ which corresponds to the halo mass as described in
Appendix\footnote{The average linear overdensity profile, equation
(\ref{overden}), holds for any value of $\nu$. 
For small $\nu$, however, random dispersion around this average
profile becomes substantial, limiting the generality of equation
(\ref{overden}).}.
The overdensity $\Delta_{0}(r)$ is
equivalent to the fractional mass perturbation $\delta M/M$ inside
radius $r$, 
\begin{equation}
\label{overden-m}
\Delta_{0}(r)=\delta M/M \propto M^{-\frac{n+3}{3}}.
\end{equation}
From equations (\ref{overmass}) and (\ref{overden-m}), we deduce that
the power-law power spectrum naturally generates a scale-free initial
condition with 
\begin{equation}
\label{epsilon-n}
\varepsilon=(n+3)/3.
\end{equation}

\begin{figure}
\includegraphics[width=84mm]{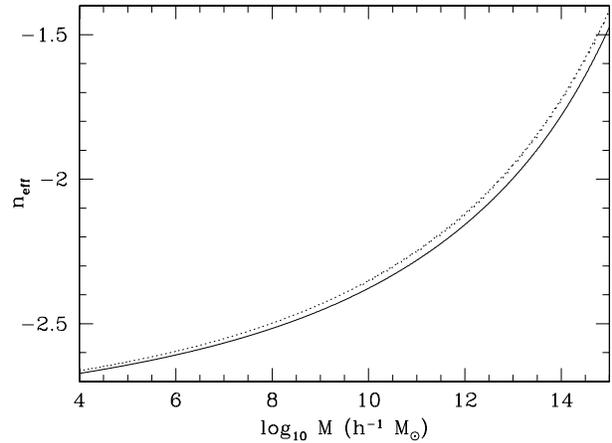}
\caption{Effective index of the power spectrum ($P(k)\propto
  k^{n_{\rm eff}}$)
vs. halo mass for the $ \Lambda $CDM universe. Appendix
describes how $n_{\rm eff}$ is calculated. The solid line is derived
  from the HS approach, which we use in this paper. The dotted line is
  derived from $\sigma_{M}$.}
\label{fig-neff}
\end{figure}

According to this model, as described in Appendix, haloes of a given
mass $M$ originate from density perturbations given by equation
(\ref{overden-m}) with $n$ determined by the primordial power
spectrum after it is transferred according to the parameters of the
background universe and the nature of the dark matter. We plot this
effective $n$ as a function of halo mass in Figure \ref{fig-neff} for
the current $\Lambda$CDM universe. The value of $n\simeq-2.5$ is a
reasonable approximation for galactic haloes
(i.e. $n\simeq-2.5\pm0.1$ for $M\simeq10^{8\pm2}M_{\sun}$, while
  $n\simeq-2.5\pm0.2$ for $M$ in the range from $10^{3}M_{\sun}$
  to $10^{11}M_{\sun}$). For haloes in the cluster mass range,
$M\sim10^{15}M_{\sun}$, $n\simeq-1.5$.

\section{Fluid approximation of dark matter dynamics}
\label{fluidapprox_sect}
We show that fluid conservation equations for a gas with 
adiabatic index $ \gamma =5/3 $
are a good approximation to the dynamics of both
CDM and SIDM haloes.
This approach has been used in the literature of stellar dynamics (e.g.
\citealt{larson, le, bett}) for the study of the gravothermal catastrophe,
where particles (stars) experience gravitational two-body
interactions. The authors integrated the Boltzmann equation with a
collision term due to gravitational two-body interactions, to obtain
a set of moment equations. They then truncated, under reasonable
assumptions, the hierarchy of such moment 
equations such that only fluid-like conservation equations
remain. The effect of gravitational two-body
interactions was naturally approximated as an effective heat conduction.
%If the number of particles of a self-gravitating system is
%small ($ \leq 10^{5} $), the gravitational two-body interaction
%cannot be fully neglected(\citealt{bt}) in dynamical timescale. 
In the CDM literature, \citet{tca} and \citet{sco} followed a similar
approach for the study of CDM haloes. Contrary to systems described by
stellar dynamics where the number of particles is small, gravitational two-body
interactions are completely negligible for CDM haloes. The authors
integrated collisionless Boltzmann equation to obtain a set of moment
equations, and truncated its hierarchy as done for stellar
dynamics. This results in fluid conservation equations for
collisionless systems.
This idea may bother some readers since, strictly speaking, the
collisionless nature of CDM prohibits the use of such an
approximation. Collisionless particles have, in
principle, an infinite set of moment equations when the Boltzmann
equation is integrated (BBGKY hierarchy; 
e.g. \citealt{bt}).
However, a couple of simple assumptions enable us to
treat CDM halo dynamics with the usual fluid conservation equations.
This section is devoted to the derivation of
the fluid approximation, which we will use in \S \ref{ss_sidm_sect} to
obtain self-similar
equations both for the collisionless CDM and the collisional SIDM
halo formation and evolution.

\citet{aas} have already applied this formalism to study CDM halo
formation. Adopting the universal mass growth history reported for CDM
N-body haloes (e.g. \citealt{wechsler}) and applying the fluid
approximation, they have shown that the resulting shape of the equilibrium halo
profile and its evolution match those of CDM N-body simulations
remarkably well. In addition to this convincing result, we shall also
justify the fluid approximation in \S \ref{fluidapprox_detail_sect}.

We develop this model under certain conditions. First, spherical
symmetry is assumed. The initial condition is given by a spherically
symmetric overdense region, and the subsequent evolution does not break
the symmetry. Second, the infall of matter is assumed to be continuous.
This can be achieved if we assume a smooth initial overdensity profile.
Third, we restrict our attention to the matter-dominated era
such that the cosmic mean density 
$ \rho _{b}\propto t^{-2} $. This condition is true in 
the $ \Lambda$CDM universe if the redshift is restricted to be
$ 1 \la z \la z_{\rm eq} $, where $z_{\rm eq}$ is the redshift for the
matter-radiation equality. 

\subsection{Fluid approximation of collisionless CDM dynamics}
\label{fluidapprox_detail_sect}
Let us describe the fluid approximation for a self-gravitating, weakly
collisional system in spherical symmetry. 
We define the average physical quantities
as follows:
\begin{equation}
\label{rho}
\rho =\int fd^{3}v,
\end{equation}
\begin{equation}
\label{aver}
\langle A\rangle \equiv \frac{\int Afd^{3}v}{\int fd^{3}v}=\frac{1}{\rho }\int Afd^{3}v,
\end{equation}
\begin{equation}
\label{velo}
u\equiv \langle v_{r}\rangle ,
\end{equation}
\begin{equation}
\label{pr}
p_{r}\equiv \rho \langle (v_{r}-\langle v_{r}\rangle )^{2}\rangle ,
\end{equation}
\begin{equation}
\label{ptheta}
p_{\theta }\equiv \rho \langle (v_{\theta }-\langle v_{\theta }\rangle )^{2}\rangle =\rho \langle v_{\theta }^{2}\rangle ,
\end{equation}
\begin{equation}
\label{pphi}
p_{\phi }\equiv \rho \langle (v_{\phi }-\langle v_{\phi }\rangle )^{2}\rangle =\rho \langle v_{\phi }^{2}\rangle ,
\end{equation}
 where $ f $ is the distribution function defined such that $
f(\textbf{r},\,\textbf{v})d^{3}r d^{3}v = $ mass within infinitesimal
volume $d^{3}r d^{3}v$ at $(\textbf{r},\,\textbf{v})$,  $ \rho  $ is the
density, $ \langle A\rangle  $ is the average value of a certain
quantity $ A $, $ u $ is the radial bulk velocity, $ p_{r} $
is the ``effective radial pressure'', and $ p_{\theta } $ is the
``effective tangential pressure''. Note that 
$\langle v_{\theta }\rangle = \langle v_{\phi }\rangle = 0$ and
$p_{\theta}=p_{\phi}$
because of spherical symmetry. Anisotropy
in the velocity dispersion occurs in general -- i.e. $ p_{r}\neq
p_{\theta } $ or anisotropy parameter $ \beta \neq 0 $ where $
\beta \equiv 1-\frac{p_{\theta}}{p_{r }} $ -- implying that we
should treat $p_{r} $ and $ p_{\theta } $ separately. In a highly
collisional system, which is well described by fluid conservation
equations, $ p_{r}=p_{\theta } $ and the usual pressure
$p=p_{r}=p_{\theta } $. 

A self-gravitating system of collisionless particles can be described by the
collisionless Boltzmann equation 
\begin{equation}
\label{boltzmann}
%\frac{df}{dt}=\left( \frac{\partial f}{\partial t}\right) _{c},
\frac{df}{dt}=0,
\end{equation}
%where $ \left( {\frac{\partial f}{\partial t}} \right) _c $ \
%is the change rate of $ f $ due to collision.
where $\frac{d}{dt}$ is the phase-space Lagrangian time-derivative,
given by
\begin{equation}
\label{total_dt}
\frac{d}{dt} = \frac{\partial}{\partial t} +
\textbf{v}\cdot\frac{\partial}{\partial \textbf{r}} 
+\textbf{a}\cdot\frac{\partial}{\partial \textbf{v}}.
\end{equation}
Throughout this paper, we will use the Newtonian approximation: when
a system is much larger than its Schwarzschild radius
and much smaller than the horizon size, motions of particles and the
temperature of the system become non-relativistic. In this limit, we
can use the non-relativistic Boltzmann transport equation.
This equation can then be written more explicitly in spherical coordinates.
For a system in spherical symmetry,
$f=f(|\textbf{r}|,\,\textbf{v})$, and equation (\ref{boltzmann}) reads
\begin{eqnarray}
%\left( \frac{\partial f}{\partial t}\right) _{c} & = & \frac{\partial
0 & = & \frac{\partial
f}{\partial t}+v_{r}\frac{\partial f}{\partial r}+\left(
\frac{v_{\theta }^{2}+v_{\phi }^{2}}{r}-\frac{\partial \Phi }{\partial
r}\right) \frac{\partial f}{\partial v_{r}} \nonumber \\ 
&  &+\frac{1}{r}\left( v_{\phi }^{2}\cot \theta -v_{r}v_{\theta
}\right) \frac{\partial f}{\partial v_{\theta }} \nonumber \\ 
&  & -\frac{v_{\phi }}{r}\left( v_{r}+v_{\theta }\cot \theta \right) \frac{\partial f}{\partial v_{\phi }}, \label{sph-boltz} 
\end{eqnarray}
where $ \Phi $ satisfies the Poisson equation
$ {\nabla}^{2} \Phi = 4 \pi G \rho $ (\citealt{bt}).
By multiplying 
equation (\ref{sph-boltz}) by $\int d^{3}v\,v_{r}^{m}v_{\theta}^{n}$, 
where $m,\,n$ are integer numbers, we can form a set of moment
equations. Moment equations from the
lowest order are
\begin{equation}
\label{mass}
\frac{\partial \rho }{\partial t}+\frac{\partial }{r^{2}\partial r}(r^{2}(\rho u))=0,
\end{equation}
\begin{equation}
\label{momentum}
\frac{\partial }{\partial t}(\rho u)+\frac{\partial }{\partial r}(p_{r}+\rho u^{2})+\frac{2}{r}(p_{r}-p_{\theta }+\rho u^{2})=-\rho \frac{Gm}{r^{2}},
\end{equation}
\begin{equation}
\label{energyr}
\rho \frac{D}{Dt}\left( \frac{p_{r}}{2\rho }\right) +p_{r}\frac{\partial u}{\partial r}=\Gamma _{1},
\end{equation}
\begin{equation}
\label{angular}
\rho \frac{D}{Dt}\left( \frac{p_{\theta }}{2\rho }\right) +\frac{p_{\theta}u}{r}=\Gamma _{2},
\end{equation}
 \[
\vdots \]
 where $ m $ is the mass enclosed by a shell at radius $ r $, 
$ \frac{D}{Dt}\equiv \frac{\partial }{\partial t}
+u\frac{\partial }{\partial r}, $
and 
\begin{eqnarray}
\label{gamma1}
%\Gamma_{1}& =& \frac{1}{2}\left. \frac{\partial }{\partial t}\right|_{c} 
%\rho \left\langle v_{r}^{2}\right\rangle +\frac{\rho
%}{r}\left\langle 2\left( v_{r}-\left\langle v_{r}\right\rangle \right)
%v_{\theta }^{2}\right\rangle \nonumber \\ 
%&  & -\frac{1}{2r^{2}}\frac{\partial }{\partial r}
%\left( r^{2}\rho \left\langle \left( v_{r}
%-\left\langle v_{r}\right\rangle \right) ^{3}\right\rangle \right) ,
\Gamma_{1}& =& 
\frac{\rho
}{r}\left\langle 2\left( v_{r}-\left\langle v_{r}\right\rangle \right)
v_{\theta }^{2}\right\rangle \nonumber \\ 
&  & -\frac{1}{2r^{2}}\frac{\partial }{\partial r}
\left( r^{2}\rho \left\langle \left( v_{r}
-\left\langle v_{r}\right\rangle \right) ^{3}\right\rangle \right) ,
\end{eqnarray}
\begin{equation}
\label{gamma2}
%\Gamma _{2}=\frac{1}{2}\left. \frac{\partial }{\partial t}\right|
%_{c}\rho \left\langle v_{\theta }^{2}\right\rangle
%-\frac{1}{4r^{4}}\frac{\partial }{\partial r}\left( r^{4}\rho
%\left\langle \left( v_{r}-\left\langle v_{r}\right\rangle \right)
%v_{\theta }^{2}\right\rangle \right) .
\Gamma _{2}=
-\frac{1}{4r^{4}}\frac{\partial }{\partial r}\left( r^{4}\rho
\left\langle \left( v_{r}-\left\langle v_{r}\right\rangle \right)
v_{\theta }^{2}\right\rangle \right) . 
\end{equation}
Equations (\ref{mass}) - (\ref{angular}) are 
conservation equations of mass, momentum,
``radial'' energy, and angular momentum respectively.
Note that equations (\ref{mass}) - (\ref{gamma2}) are all in exact
form, and the hierarchy of equations is not closed in principle. 

\begin{figure}
\includegraphics[width=84mm]{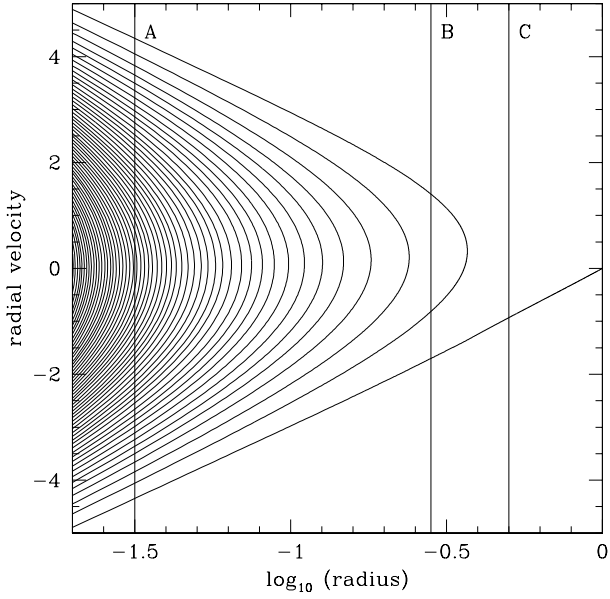}
\caption{Self-similar collisionless halo formation for
  $\varepsilon=1$: Phase-space diagram of \citet{bert} solution as an
illustration of the ``skewless distribution'' assumption. Along
the line A there are a large number of shell-crossings, which are almost
symmetric around $ \langle v_{r} \rangle $, and along the line C, as
there is only one stream line, it is intrinsically skewless. On the
contrary, there are only three stream lines along the line B, so the
assumption is not good in the region inside and close to the outermost
caustics. Tangential velocity distribution would have a similar
  behaviour when tangential motion is allowed.}  
\label{fig-phase}
\end{figure}

Now we make a further simplification that the distribution of $v_{r}$ is
skewless -- $v_{\theta}$ and $v_{\phi}$ are 
naturally skewless because of
spherical symmetry. In other words, we assume that $v_{r}$ has
a symmetric distribution around $\langle v_{r} \rangle$. It is not
straightforward to show that $\Gamma_{1}$ and $ \Gamma_{2}$ are
negligible in equations (\ref{energyr}) and (\ref{angular}). However,
we demonstrate that the assumption of ``skewlessness'' in the fluid
approximation yields results which are in good agreement with the
purely collisionless CDM structure, for specific
examples\footnote{\citet{sco} argue that an initially
skewless distribution function $f$ will remain skewless even after
evolution. It is true if one is 
interested in the fine-grained distribution, or each stream line. However,
if one considers coarse-grained quantities as defined by equation
(\ref{aver}), the contribution from multiple stream lines should be
counted and the overall distribution is not skewless, in general.}.   
As shown in the \citet{bert} solution, for instance, collisionless
particles (shells) form a quasi-symmetric winding structure in the phase
space as shown in Figure \ref{fig-phase}. This indicates
that we may neglect terms
which arise as a result of skewness.
Equations (\ref{mass}) - (\ref{energyr}) 
with the condition $p_{\theta}=0$ and $\Gamma_{1}=\Gamma_{2}=0$, 
\begin{equation}
\label{mass-bert}
\frac{\partial \rho }{\partial t}+\frac{\partial }{r^{2}\partial r}(r^{2}(\rho u))=0,
\end{equation}
\begin{equation}
\label{momentum-bert}
\frac{\partial }{\partial t}(\rho u)+\frac{\partial }{\partial r}(p_{r}+\rho u^{2})+\frac{2}{r}(p_{r}+\rho u^{2})=-\rho \frac{Gm}{r^{2}},
\end{equation}
\begin{equation}
\label{energyr-bert}
\rho \frac{D}{Dt}\left( \frac{p_{r}}{2\rho }\right) +p_{r}\frac{\partial u}{\partial r}=0,
\end{equation}
can be used to solve purely radial problems, such as the
spherical infall problems with similarity solutions by
\citet[$\varepsilon=1$]{bert} and \citet[with $\varepsilon=1/6$]{fg}.
As seen in Figure \ref{fig-bert}, the solution
to equations (\ref{mass-bert}) - (\ref{energyr-bert}) is in good
agreement with the true solution. We have also found, as seen in
Figure \ref{fig-e16}, an excellent agreement between the collisionless
solution and the one produced by a radial fluid approximation in the case of
$\varepsilon=1/6$ (the $\varepsilon=1/6$ case is of main
interest in this paper). 
The difference observed at caustics -- places where the density
becomes infinite -- is negligible, because caustics do not affect the
overall dynamics of the halo. Since the skew-free assumption naturally
neglects dynamically unimportant structure (e.g. caustics) while
accurately reproducing the profile of the exact solution in these
radial cases, it may also be applied to describe CDM haloes, in which
particles have a tangential motion as well.

The final assumption is that 
inside the virialized structure (i.e. postshock region) the velocity
dispersion is isotropic, or $p_{r}=p_{\theta }$. 
This is an empirical assumption: CDM haloes in cosmological N-body
simulations show mild anisotropy. 
For instance,
\citet{carlberg} show that CDM haloes in their numerical simulation can
be well-fitted by a fitting formula
\begin{equation}
\label{anisotropy}
\beta(r)=\beta_{m}\frac{4r}{r^{2}+4},
\end{equation}
where $r$ is in units of $r_{200}$, and $\beta_{m}=[0.3-0.5]$
(see also, e.g., \citealt{thomas} and \citealt{ckk}).
As we
will show in the following sections, our similarity solutions have
a virial radius $r_{564}\simeq 0.6 \,r_{200}$, which then results in
the maximum anisotropy $\beta\sim[0.17 - 0.28]$. This fact enables us to use
$\beta=0$ to a good approximation.

With these assumptions (spherical symmetry, skew-free velocity
distribution, and isotropic velocity dispersion),
the usual fluid conservation equations are obtained.
They are
\begin{equation}
\label{mass-cdm}
\frac{\partial \rho }{\partial t}+\frac{\partial }{r^{2}\partial r}(r^{2}(\rho u))=0,
\end{equation}
\begin{equation}
\label{momentum-cdm}
\frac{\partial }{\partial t}(\rho u)+\frac{\partial }{\partial r}(p+\rho u^{2})+\frac{2}{r}\rho u^{2}=-\rho \frac{Gm}{r^{2}},
\end{equation}
\begin{equation}
\label{energy-cdm}
\frac{D}{Dt}(\frac{3p}{2\rho })=-\frac{p}{\rho }\frac{\partial }{r^{2}\partial r}(r^{2}u),
\end{equation}
which are identical to the fluid conservation equations for a $\gamma=5/3$
gas in spherical symmetry.
Resemblance of these equations to fluid equations indicate that we
can expect an effective ``shock'' even for a collisionless system
because of the hyperbolicity of these equations. This is also illustrated
in Figure \ref{fig-phase}. For instance, the density jump occurs when
one moves from the ``pre-shock'' region (line C; one stream line) to the
``post-shock'' region (line B; three stream lines). However, one should
be careful in using 
this formalism because the approximation becomes worse where there are
only a small number of phase-space windings.

\subsection{Fluid approximation of SIDM haloes}
\label{fluidapprox_sidm_sect}
We now consider the effect of SIDM collisionality. We adopt a simple,
heuristic approach to account for the heat conduction as a result of
finite cross-section.
This will change the energy conservation
equation (equ. [\ref{energy-cdm}]) to
\begin{equation}
\label{energy-sidm}
\frac{D}{Dt}(\frac{3p}{2\rho })=-\frac{p}{\rho }\frac{\partial
}{r^{2}\partial r}(r^{2}u)-\nabla \cdot \textbf{f},
\end{equation}
where $\textbf{f}$ is the heat flux.

The heat flux resulting from SIDM interaction has been derived by
BSI. We briefly describe its derivation. When the scattering
mean free path is much smaller than the system size (short mean free
path limit, or diffusion limit), the heat flux (equivalent to
$\frac{L}{4\pi r^{2}}$ in BSI) reads
\begin{equation}
\label{f-short}
{\textbf{f}_{\rm smfp}}=-\frac{3b}{2\sigma} \sqrt{\frac{p}{\rho }}
\frac{\partial }{\partial r}\left( \frac{p}{\rho }\right) \hat{r},
\end{equation}
where $b$ is an effective impact parameter of order unity, which is
1.002 for elastic scattering of hard spheres.
On the contrary, when the mean free path is much larger than the
system size (long mean free path limit), the proper length scale of
heat transfer is not the mean 
free path (a multiple of the system size) but the system size, while
the time scale is still the 
relaxation time. This results in
\begin{equation}
\label{f-long}
{\textbf{f}_{\rm lmfp}}=-\frac{3}{2} a b \sigma \sqrt{\frac{p}{\rho }}
\frac{p}{4\pi G}
\frac{\partial }{\partial r}\left( \frac{p}{\rho }\right) \hat{r},
\end{equation}
where $a=2.26$ for elastic scattering of hard spheres. 

In the end, we use a hybrid expression which is roughly valid in the
intermediate regime as well as in two extreme regimes (short mean free
path limit and long mean free path limit)\footnote{Note that the
intermediate regime described by equation~(\ref{flux}) is indeed a
mere interpolation of two different regimes. We 
believe, however, that this is a good approximation: the intermediate
value should not be too different from this smooth and continuous
interpolation of two regimes.
},
\begin{equation}
\label{flux}
{\textbf {f}}=-\frac{3}{2}ab\sigma \sqrt{\frac{p}{\rho }}\left( a\sigma ^{2}+\frac{4\pi G}{p}\right) ^{-1}\frac{\partial }{\partial r}\left( \frac{p}{\rho }\right) \hat{r}.
\end{equation}
Throughout this paper, we adopt $a=2.26$ and $b=1.002$.

\subsection{Shock jump conditions}
\label{shockjump_sect}
As we have shown in previous sections, fluid conservation
equations can be used and we expect an accretion ``shock'' to
occur. Across the shock, the matter, momentum and energy fluxes
should all be continuous. Mathematically, a flux is a term acted upon by 
$\frac{\partial}{\partial r}$ in the conservation equations expressed in
Eulerian coordinates.

In the adiabatic case\footnote{We will use the term ``adiabatic'' for
the case where there is no heating mechanism (e.g. conductive
heating due to elastic scattering) other than the shock heating.},
therefore, we obtain the usual adiabatic shock 
jump conditions from equations (\ref{mass-cdm}) - (\ref{energy-cdm}):
\begin{equation}
\label{jump-mass-cdm}
\left[ \rho \bar{u} \right]=0,
\end{equation}
\begin{equation}
\label{jump-momentum-cdm}
\left[ p + \rho \bar{u}^2 \right] = 0,
\end{equation}
\begin{equation}
\label{jump-energy-cdm}
\left[ \rho \bar{u} \left( \frac{3p}{2\rho}+\frac{1}{2} \bar{u}^2 +
\frac{p}{\rho} \right)
\right] = 0,
\end{equation}
where $[A]\equiv A(\textrm{preshock}) - A(\textrm{postshock}) $,
$\bar{u}$ ($= u -u_{\rm s}$) is the bulk radial velocity in
the shock frame,
and equation (\ref{jump-energy-cdm}) comes from 
equation (\ref{energy-cdm}) if expressed in Eulerian coordinates by converting
$\frac{D}{Dt}$ into $\frac{\partial}{\partial t}
+ u \frac{\partial}{\partial r} $.

For an SIDM case, as the heat conduction is included, the energy jump
condition will instead be
\begin{equation}
\label{jump-energy-sidm}
\left[ \rho \bar{u} \left( \frac{3p}{2\rho}+\frac{1}{2} \bar{u}^2 +
 \frac{p}{\rho}  + f \right)\right] = 0,
\end{equation}
where $ \textbf{f}= f \hat{r} $.

\section{Self-Similar Model for SIDM Haloes in the Matter-Dominated Era}
\label{ss_sidm_sect}
In this section we first show that formation and evolution of SIDM
haloes in galactic scale are well approximated by self-similar equations.
We then apply the fluid approximation to the
problem and derive a set of ordinary differential equations. We will
explain in detail how to solve these equations with proper
boundary conditions.

\subsection{Self-similarity of SIDM haloes}
\label{ss_sidm_small_sect}
In \S \ref{hs_sect}, we described how a
self-similar collapse model fits in cosmology by relating the
logarithmic slope of the power spectrum, $n$, to the initial linear
overdensity profile parametrized by $\varepsilon$. We, therefore,
should first know what $n$ is relevant to the problem we
solve.

When collisionality of particles enters the system, its corresponding
length scale comes into play, which in general does not grow in proportion
to $ r_{\rm ta} $. Similarly, a new time scale will enter the system.
However, we can still make the system self-similar after the addition
of collisionality by some fine-tuning. The conductive heating
term $ \nabla \cdot {\textbf {f}} $ enters the energy equation,
and it should have the same
time dependence as does the adiabatic change
of the thermal energy, $ \rho \frac{d}{dt}\left( \frac{3p}{2\rho}
\right) $.  Noting that $ \rho \frac{d}{dt} \left( \frac{3p}{2\rho}
\right) \propto r_{\rm ta}^{2}t^{-5} $ 
and $ \nabla \cdot {\textbf {f}}\propto r_{\rm ta}^{3}t^{-7} $, the
condition 
\begin{equation}
\label{condition}
\xi =2,\,\,\varepsilon =\frac{1}{6}
\end{equation}
preserves self-similarity of the system. This condition,
$\varepsilon=1/6$, is equivalent to $n=-2.5$, which can be seen
from equation (\ref{epsilon-n}). The turnaround radius and mass grow
in time as
\begin{equation}
\label{growth}
r_{\rm ta}\propto t^{2}, \,\,\, M\propto t^{4}.
\end{equation}

Now we find an intriguing coincidence. As described in \S \ref{hs_sect},
$n=-2.5$ is a good approximate value for haloes of galactic mass.
SIDM haloes in galactic scales are, therefore, well described by
self-similar equations. The following 
sections are dedicated to the detailed description of SIDM similarity
solutions.

\subsection{Basic equations and problem solving scheme}
\label{solve_sect}
Under the condition of self-similarity, one can convert seemingly
time-dependent equations into ordinary differential equations by
properly scaling physical parameters. The turnaround radius $r_{\rm ta}$
is a natural choice for a length scale. With time $t$ and the
turnaround radius $r_{\rm ta}$, we define dimensionless physical
quantities -- radius, velocity, density, pressure, mass and heat flux
-- as follows: 
\begin{equation}
\label{rr}
\lambda =r/r_{\rm ta},
\end{equation}
\begin{equation}
\label{vv}
V(\lambda )=v/\left( \frac{r_{\rm ta}}{t}\right) ,
\end{equation}
\begin{equation}
\label{rhorho}
D(\lambda )=\rho /\rho _{b},
\end{equation}
\begin{equation}
\label{pp}
P(\lambda )=p/\left[ \rho _{b}\left( \frac{r_{\rm ta}}{t}\right) ^{2}\right] ,
\end{equation}
\begin{equation}
\label{mm}
M(\lambda )=m/\left( \frac{4\pi }{3}\rho _{b}r_{\rm ta}^{3}\right) ,
\end{equation}
\begin{equation}
\label{fluxflux}
F(\lambda) = f / \left[ \rho_{b} \left( \frac{r_{\rm ta}}{t} \right)^3 \right].
\end{equation}

Collapsing shells, which are assumed to be cold initially, obey
Newton's law
\begin{equation}
\label{newton}
\frac{d^2 r}{d t^2}=-\frac{Gm}{r^{2}}
\end{equation}
where $m$ is the mass enclosed by radius $r$. For
the initial density perturbation defined by equation (\ref{overmass}), 
this Newtonian motion can be described by a set of
parametric equations, as follows (\citealt{abn}; see also
\citealt[][for case of $\varepsilon=1$]{bert}):
\begin{equation}
\label{pre-rr}
\lambda =\sin ^{2}(\theta /2)\, \left( \theta -\sin \theta \over \pi \right) ^{-\xi },
\end{equation}
\begin{equation}
\label{pre-vv}
V(\lambda )=\lambda \frac{\sin \theta (\theta -\sin \theta )}{(1-\cos \theta )^{2}},
\end{equation}
\begin{equation}
\label{pre-rhorho}
D(\lambda )=\frac{9}{2}\frac{(\theta -\sin \theta )^{2}}{(1-\cos \theta )^{3}(1+3\epsilon \chi )},
\end{equation}
\begin{equation}
\label{pre-mm}
M(\lambda )=\lambda ^{3}\frac{9}{2}\frac{(\theta -\sin \theta )^{2}}{(1-\cos \theta )^{3}},
\end{equation}
 where $ \chi =1-(3/2)(V(\lambda )/\lambda )$. Equations
 (\ref{pre-rr}) - (\ref{pre-mm}) therefore describe preshock motion.

The postshock motion of shells is described by full hydrodynamic equations.
Equations (\ref{mass-cdm}), (\ref{momentum-cdm}), (\ref{energy-sidm}),
(\ref{flux}) and the definition
of the infinitesimal mass $ dm=4\pi r^{2}dr $ can be written
with these dimensionless quantities as a set of ordinary differential
equations:\begin{equation}
\label{mmass}
(V-2 \lambda )D'+DV'+\frac{2DV}{\lambda }-2D=0,
\end{equation}
\begin{equation}
\label{mmomentum}
(V-2 \lambda )V'+V=-\frac{P'}{D}-\frac{2}{9}\frac{M}{\lambda ^{2}},
\end{equation}
\begin{equation}
\label{eenergy}
(V-2\lambda )\left( \frac{P'}{P}-\frac{5D'}{3D}\right) =-\frac{10}{3}
-\frac{2(F \lambda^{2})'}{3P\lambda ^{2}}. 
\end{equation}
\begin{equation}
\label{fflux}
F=-\frac{3ab}{2Q'}(a+\frac{2}{3Q'^{2}P})^{-1}\sqrt{\frac{P}{D}}\frac{d}{d\lambda }\left( \frac{P}{D}\right) ,
\end{equation}
\begin{equation}
\label{dmdm}
M'=3\lambda ^{2}D,
\end{equation}
where the prime indicates differentiation with respect to $ \lambda $,
and the nondimensional collisionality parameter $ Q' $ is defined as
$ Q'\equiv \sigma \rho _{b}r_{\rm ta} $. We will later use a nondimensional constant $ Q\equiv \sigma \rho _{b}r_{\rm s}=\lambda _{\rm s}Q', $
which is more directly related to the collision rate of an SIDM particle
in a virialized structure with the shock radius $ r_{\rm s}. $ 
For instance, the number of collisions a particle experiences in a
time $ \Delta t $ 
is given by $ N\equiv \sigma \rho v_{\rm rel}\Delta t $
(e.g. \citealt{bur2}). The conversion of $Q$ into $N$ is
straightforward:
\begin{equation}
\label{qn}
N=\frac{\rho}{\rho_{b}}\frac{v_{\rm rel}\Delta t}{r_{\rm s}} Q 
 =a\sqrt{\frac{p}{\rho}}\frac{\rho}{\rho_{b}}\frac{\Delta t}{r_{\rm s}} Q.
\end{equation}
Here we used the relation $v_{\rm rel}= a \sqrt{p/\rho}$,
where $a=2.26$ again, which relates the average thermal velocity to
the relative velocity for particles interacting elastically as hard spheres
(equs. [7.10.13], [12.2.12] in \citealt{reif}).

Different solutions arise for different values of $ Q(Q') $. To solve
the coupled ordinary differential equations (equs. [\ref{mmass}] -
 [\ref{dmdm}]), 
we need to connect the preshock values given by equations
(\ref{pre-rr}) - (\ref{pre-mm} with $ \xi =2$) to the postshock
values. This is described by 
the shock jump conditions (equs. [\ref{jump-mass-cdm}],
[\ref{jump-momentum-cdm}], and [\ref{jump-energy-sidm}]) and the
continuity of mass, which are
expressed 
by nondimensional variables as
\begin{equation}
\label{j-rho}
D_{2}=\left( 1-\frac{2F_{2}}{P_{2}(V_{1}-2 \lambda _{\rm s})}\right) ^{-1}4D_{1},
\end{equation}
\begin{equation}
\label{j-p}
P_{2}=\left( \frac{3}{4}+\frac{F_{2}}{2 P_{2}(V_{1}-2 \lambda _{\rm s})}\right) D_{1}(V_{1}-2 \lambda _{\rm s})^{2},
\end{equation}
\begin{equation}
\label{j-v}
V_{2}=2 \lambda _{\rm s}+\left( 1-\frac{2F_{2}}{P_{2}(V_{1}-2 \lambda _{\rm s})}\right) \frac{1}{4}(V_{1}-2 \lambda _{\rm s}),
\end{equation}
\begin{equation}
\label{j-m}
M_{2}=M_{1},
\end{equation}
where the subscript 1 denotes preshock values, while 2 denotes postshock
values. Note that $P_{1}=0 $ and $F_{1}=0$ because we assume cold infall.
Note also that without terms containing $ F_{2}, $ equations
(\ref{j-rho}) - (\ref{j-m}) are identical to the adiabatic jump conditions
for $ \gamma =5/3 $ gas (see equs. [6a] - [6d] in \citealt{abn}).
These additional terms arise because of finite conductivity in the
postshock region. Finally, the
inner boundary conditions are\begin{equation}
\label{i-m}
M(\lambda =0)=0,
\end{equation}
\begin{equation}
\label{i-v}
V(\lambda =0)=0,
\end{equation}
and
\begin{equation}
\label{i-f}
F(\lambda =0)=0.
\end{equation}

In principle, the fluid equations (equs. [\ref{mmass}] - [\ref{dmdm}]),
preshock equations (equs. [\ref{pre-rr}] - [\ref{pre-mm}]), jump
conditions (equs. [\ref{j-rho}] - [\ref{j-m}]), and inner boundary
conditions (equs. [\ref{i-m}] - [\ref{i-f}]) yield a unique solution.
This solution was obtained numerically, by iteration, as follows. We
arbitrarily chose the shock location $\lambda_s$ (close to that of the
adiabatic solution), central density $D(0)$ and central pressure $P(0)$.
We then integrated differential equations from $\lambda=0$ outward,
using the LSODE (Livermore Solver for Ordinary Differential Equations), in
which case variables 
behaved well. At the chosen shock location, we then obtained
preshock values using the jump condition. If these values differed
from those obtained from equations (\ref{pre-rr}) - (\ref{pre-mm}), we
went back and chose a different $\lambda_s$, $D(0)$, and
$P(0)$. We iterated this process until the jump condition was
satisfied to a given error tolerance. This way we could get approximate
solutions 
for several selected values of $ Q $ (listed in Table \ref{solution}),
all of which satisfied the shock jump conditions (equs. [\ref{j-rho}]
- [\ref{j-m}]) to less than 0.1\% error. This was
a very tedious and time-consuming job, but automation of the iterations
made it possible to achieve this goal.

In practice we could not perform integration from $ \lambda =0 $
because the differential 
equations have a coordinate singularity at the centre. Instead, we
performed the integration from some small $\lambda$, using
asymptotic forms for $\lambda \ll 1$: $D \simeq D(0)$, 
$P \simeq P(0)$, 
$V \simeq \frac{2}{3} \lambda$, 
$M \simeq D(0) \lambda^3$, 
and $F \simeq -\frac{5}{3} P(0) \lambda$.

\section{Results}
\label{result_sect}
\begin{figure}
\includegraphics[width=84mm]{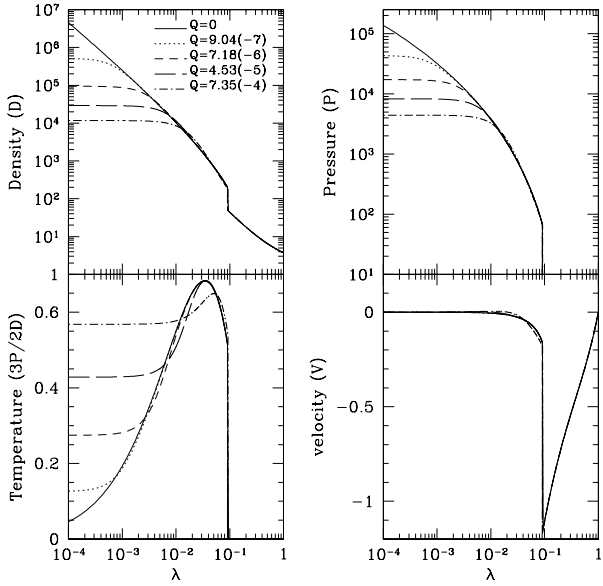}
\caption{Similarity solution dimensionless profiles for low-$ Q $
  regime. $Q=0$ means ``no conduction,'' i.e. ``adiabatic'' post-shock gas.
As $ Q $ increases, core density decreases and core temperature
increases. Dimensionless similarity variables follow the definitions
in \citet{bert}.}
\label{fig-low}
\end{figure}

\begin{figure}
\includegraphics[width=84mm]{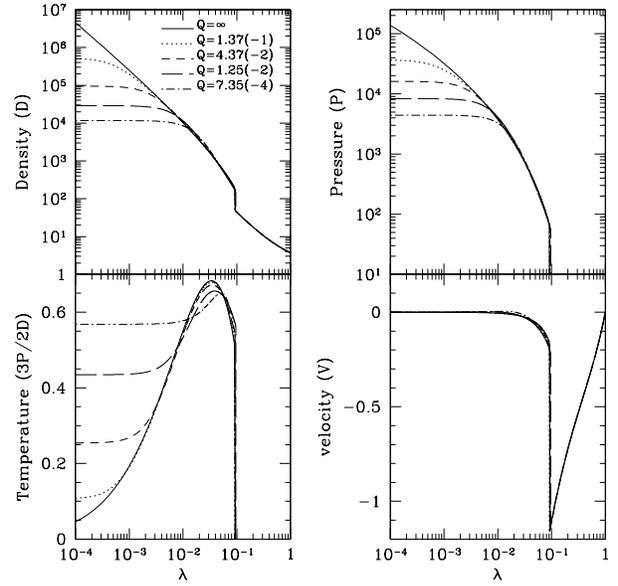}
\caption{Similarity solution dimensionless profiles for the high-$ Q
  $ regime. 
Profiles are indistinguishable from those in Figure \ref{fig-low}, even
though the $Q$ values are quite different. The effect of $Q$ is
reversed compared to the low-$Q$ regime: as $ Q $ increases, core
  density increases and core temperature 
decreases.}
\label{fig-high}
\end{figure}

We solved the coupled set of ordinary differential equations which
yield the
similarity solutions for SIDM haloes for the full range of 
values of the dimensionless collisionality parameter, $ Q $.
The results are plotted in Figures \ref{fig-low} and
\ref{fig-high}. Before we discuss the solutions for SIDM haloes with
conduction ($Q \neq 0$), we begin with a description of the solution
for $Q=0$, the case without conduction. As we shall see, the latter
yields a density profile which is in good agreement with the results
of N-body simulations of haloes in the standard CDM model. We are
justified, therefore, in applying the same self-similar infall model
with $Q\neq 0$ to describe SIDM haloes.

Let us briefly describe the general properties of the SIDM similarity
solutions and emphasize the importance of cosmological infall, before
we describe the results in full detail.

\subsection{Self-similar haloes without conduction($\varepsilon=1/6$):
  an analytical model for CDM N-body results}
\label{ssCDM_sect}
\begin{figure}
\includegraphics[width=84mm]{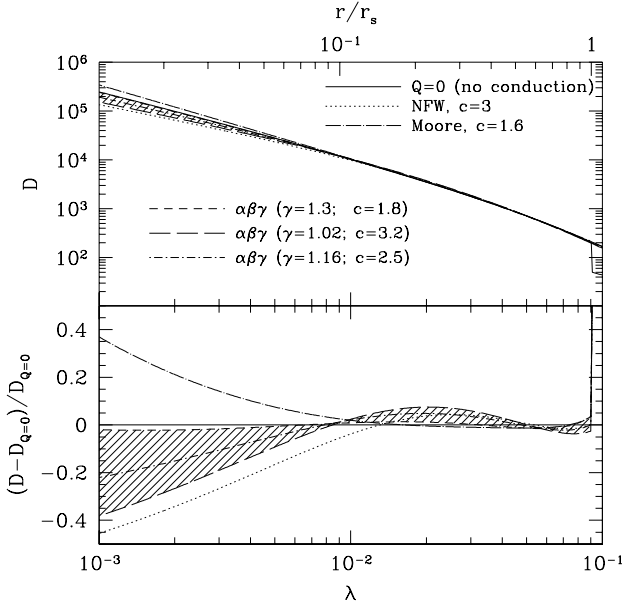}
\caption{Comparison of self-similar halo profile without conduction
  with N-body results for CDM haloes: (top panel) Density (in units of
  cosmic mean density) vs. radius (in units of current turnaround
  radius) for similarity solution ($Q=0$; $\varepsilon=1/6$) (solid),
  the best-fitting NFW profile ($c=3$), Moore profile ($c=1.6$) and
  several $\alpha \beta 
  \gamma$ profiles ($\alpha=1$, $\beta=3$) for different values of
  $\gamma$, $\gamma=1.3$ (best fit), 1.02, 1.16, as
  labelled. \citet{dms} find N-body results for CDM haloes best fit by
  ($\alpha$, $\beta$, $\gamma$) = (1, 3, $\gamma=1.16\pm 0.14$)
  (shaded region); (bottom panel) fractional deviation of the N-body
  results from similarity solution. Note that $\lambda = \lambda_{\rm s}
  \simeq 0.09$ corresponds to $r_{\rm s}=r_{564}\simeq 0.6 \, r_{200}$.
}
\label{fig-adia}
\end{figure}

Before treating soft-core solutions, we describe properties of the adiabatic
solution for $ \varepsilon = 1/6 $.
As we shall show in what follows,
the adiabatic infall solution for $ \varepsilon =1/6 $ case
resembles standard CDM haloes in many respects. First, it has a density cusp with a logarithmic
slope $\simeq-1.27$ for $4\times 10^{-3} < r/r_{200} <1.4\times
10^{-2}$, where $ r_{200}$ is the radius in which the average
density is $200\rho_{b}$, if the density is extrapolated beyond
$r_{\rm s}$ with the NFW profile that best-fits this adiabatic solution
(Fig. \ref{fig-adia}).
Note that it is possible to have a slope shallower than
$-2$ because particle velocities are allowed to have a tangential component.
If no tangential motion is allowed, the value $ -2 $ is the shallowest
slope possible for collisionless dark matter 
haloes (\citealt{rt}; \citealt{tca}; \citealt{bert-review}).

Second, the temperature profile is very similar to that of CDM haloes.
The temperature is zero at the centre, rises to a maximum at some $
\lambda $, 
and then falls to a nonzero value at the shock. The most important part
is that the temperature inversion exists. If not, the addition of conductivity
will worsen the situation by initiating gravothermal catastrophe,
rather than generating a soft core (\citealt{cavf}).

The average density inside the shock radius
is found to be $ 564\rho _{b} $ for the adiabatic solution.
This value is larger than the average
density ($ \simeq 200 $) usually adopted by convention to identify
virialized CDM haloes in N-body simulations. In terms of  
radius, $r_{\rm s}=r_{564}\simeq 0.6 r_{200}$. 
Our similarity solution, therefore, yields a shock at a smaller value of
radius than is typically used to characterize the virial radius of CDM
haloes in N-body simulations.
However, as long as we focus on the
properties inside $ r_{564} $, the adiabatic solution is a good
fit to CDM haloes as we shall see below. 
As seen in Figure \ref{fig-adia}, except for the innermost central region
where the ambiguity of density slope exists between $ -1$ \citep{nfw}
and $ -1.5$ \citep{moore}, our adiabatic solution agrees with
the NFW profile and the Moore profile to within $ 10\% $,
depending on the concentration parameter. 
Compared to halo profiles studied by \citet{dms}, the agreement is
even better (Fig. [\ref{fig-adia}]). We also find that a local
logarithmic slope slowly changes to shallower values as one
approaches the centre, which agrees with the trend reported by \citet{netal}.
We demonstrate this as follows.

We compare the adiabatic solution with the CDM N-body halo profiles
mentioned above: i.e. the NFW profile,
\begin{equation}
\label{nfw}
\rho _{\rm NFW}=\frac{\rho _{\rm sc}}{\left( r/r_{\rm sc}\right) \left( 1+r/r_{\rm sc}\right) ^{2}},
\end{equation}
the Moore profile,
\begin{equation}
\label{moore}
\rho _{\rm M}=\frac{\rho _{\rm sc}}{\left( r/r_{\rm sc}\right) ^{1.5}\left( 1+\left( r/r_{\rm sc}\right) ^{1.5}\right) },
\end{equation}
and the $\alpha \beta \gamma$ profile (e.g. \citealt{dms}),
\begin{equation}
\label{abc}
\rho_{\alpha \beta \gamma} = \frac{\rho_{\rm sc}}{\left(
  r/r_{\rm sc}\right)^{\gamma} \left(
  1+(r/r_{\rm sc})^{\alpha}\right)^{(\beta-\gamma)/\alpha}},
\end{equation}
by finding the best-fitting parameters. 
The NFW and Moore profiles have two free parameters ($\rho_{\rm sc}$ and
$r_{\rm sc}$), while the $\alpha \beta \gamma$ profile has three
($\rho_{\rm sc}$, $r_{\rm sc}$, and $\gamma$) when $\alpha$ and $\beta$ are
fixed as in \citet{dms}. We depict this comparison in Figure
\ref{fig-adia}, with the 
``concentration parameter'' -- defined by $c \equiv r_{200}/r_{\rm sc}$ -- 
used to find the best-fitting N-body halo profiles. For $\alpha \beta
\gamma$ profiles, we use the best-fitting profiles by \citet{dms}, namely
$\alpha = 1$, $\beta=3$, $\gamma = 1.16 \pm 0.14$. Among these
profiles, the $\alpha \beta \gamma$ profile with $\alpha=1$, $\beta=3$
and $\gamma=1.3$ provides the best agreement with the adiabatic solution.

\begin{figure}
\includegraphics[width=84mm]{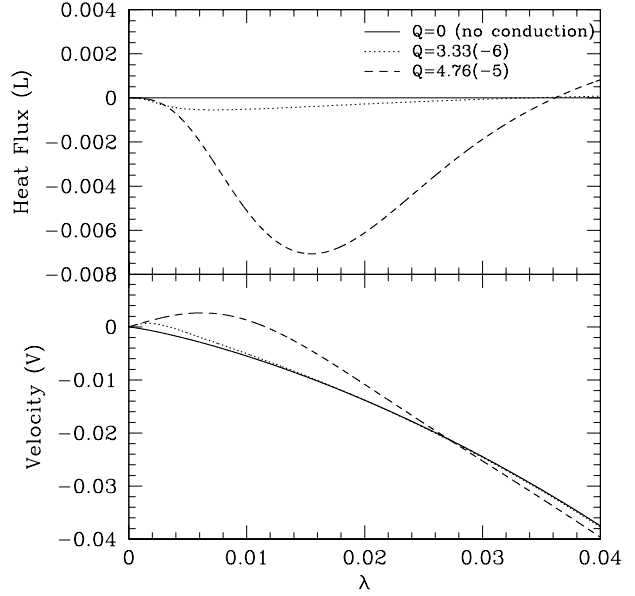}
\caption{Profiles of heat flux and bulk velocity. 
Negative heat flux in the 
core indicates that heat is transferred from the halo to the core. The
positive velocity bump indicates that the central density keeps being
flattened by a net expansion of the core. Two different $Q$ values
were selected for comparison.}
\label{fig-simb}
\end{figure}

The relatively low concentration parameter, $c\simeq 3$, required
for the best-fitting NFW profile deserves attention with respect to
the cosmological mass accretion rate.
This value is a bit small compared to the
typical concentration parameters, $c\sim[4-20]$, observed at $z=0$
in N-body simulations. 
Recent high resolution N-body simulation results, however,
report such low concentration parameters if CDM haloes are observed at
higher redshifts (e.g. \citealt{tkgk}). 
Individual haloes in CDM N-body simulations evolve over time, on average,
through a continuous sequence of universal-shaped mass profiles of
increasing total mass (\citealt{tkgk}; \citealt{vandenbosch};
\citealt{wechsler}). 
This Lagrangian mass evolution can be characterized by a universal
mass accretion history:
\begin{equation}
\label{wechsler-mass}
M(a)=M_\infty \exp (-2a_{f}/a),
\end{equation}
where $a$ is the cosmic scale factor and $a_{\rm f}$ 
is some particular value of $a$, such as that at which $d\log M
/d\log a=2$ (\citealt{wechsler}). 
As the mass of each halo grows with time due to
the average effect of mergers and smooth infall, so does the concentration
parameter $c$ of its density profile, roughly as $c(a)/c(a_{\rm
  f})\propto a/a_{\rm f}$ 
for $a/a_{\rm f}>1$ (\citealt{wechsler}), after hovering at low
values $c \approx 
2-4$ during the initial phase of most rapid mass assembly prior to
$a_{\rm f}$ (\citealt{tkgk}).
Our similarity solution has $\frac{d\log M}{d\log a}=6$, as seen in
equation (\ref{growth}), and this corresponds to $a=a_{f}/3$. As our
solution corresponds to a very early epoch in the halo formation history,
or a fast accretion rate, such a low concentration parameter is a
natural outcome.

In summary, we have shown that the adiabatic solution for
$\varepsilon = 1/6$ approximates the N-body CDM halo profiles well,
thus providing physical insights about their origin. In the following
sections, we will describe SIDM halo solutions which arise as a
result of non-zero conductivity.

\subsection{Self-similar haloes with conduction($\varepsilon=1/6$):
 an analytical Model for SIDM haloes}
\subsubsection{Low-$Q$ regime}
\label{lowq_sect}
We define the low-$ Q $ regime as $ Q\leq Q_{\rm th}=7.35\times 10^{-4} $.
All the solutions have an isothermal,
flat-density core, except for extremely small $ Q $,
where the system undergoes adiabatic collapse
just as it would in the absence of SIDM conductivity.
In this regime, as $ Q $ increases, the core density and pressure 
decrease. 
In other words, higher SIDM collisionality corresponds to a flatter
core. This trend is also observed in the temperature profile. As $Q$
increases, heat is more effectively transferred into the centre to
equalize the temperature (Fig. [\ref{fig-simb}]). Dependence of the central density,
temperature and the location of the shock on $Q$ is listed in Table
\ref{solution} (see also Fig. [\ref{fig-low}]).

This regime roughly corresponds to the long mean
free path limit. In this case, the heat flux is an increasing function
of the SIDM cross section $\sigma$ (eq. [\ref{f-long}]). 
As an increase in $Q$
is achieved by an increase in $\sigma$, the heat flux also increases
correspondingly. In \S \ref{why_sect}, a more detailed description
will be given about the quantitative relation between the
low(high)-$Q$ regime to the long(short) mean free path limit.

In \S \ref{nbody_sidm_sect}, we will also show that cosmological SIDM N-body
simulations to date have been performed only in the low-$Q$
regime. They see a monotonic behaviour of the halo profile depending on
$\sigma$ -- as $\sigma$ increases, core density decreases. In the
next section, we will show that there is an additional regime, the
high-$Q$ regime,
where this behaviour is reversed. 

\begin{table}
\caption{Central parameters and the shock location for different $Q$
  solutions. This table lists the whole range of $Q$.}
\label{solution}
\begin{tabular}{ccccc}
\hline 
$ Q $&
$ Q/Q_{\rm th} $&
$ D(0) $&
$ \frac{3P(0)}{2D(0)} $&
$ \lambda _{\rm s} $\\
\hline
$ 0 $&
$ 0 $&
$ \infty  $&
$ 0 $&
$ 9.0434 (-2) $\\
$ 9.04 (-7) $&
$ 1.23 (-3) $&
$ 5.199 (5) $&
$ 1.258 (-1) $&
$ 9.0434 (-2) $\\
$ 7.18 (-6) $&
$ 9.77 (-3) $&
$ 9.426 (4) $&
$ 2.747 (-1) $&
$ 9.0434 (-2) $\\
$ 4.53 (-5) $&
$ 6.16 (-2)  $&
$ 2.914 (4) $&
$ 4.289 (-1) $&
$ 9.060 (-2) $\\
$ 7.35 (-4) $&
$ 1 $&
$ 1.169 (4) $&
$ 5.681 (-1) $&
$ 9.260 (-2) $\\
$ 1.25 (-2) $&
$ 1.70 (1) $&
$ 2.882 (4) $&
$ 4.346 (-1) $&
$ 9.623 (-2) $\\
$ 4.37 (-2) $&
$ 5.95 (1) $&
$ 9.515 (4) $&
$ 2.540 (-1) $&
$ 9.294 (-2) $\\
$ 1.37 (-1) $&
$ 1.86 (2) $&
$ 5.199 (5) $&
$ 1.062 (-1) $&
$ 9.133 (-2) $\\
%$ 7.19 (-1) $&
%$ 9.78 (2) $&
%$ 1.399 (7) $&
%$ 1.615 (-2) $&
%$ 9.060 (-2) $\\
$ \infty  $&
$ \infty  $&
$ \infty  $&
$ 0 $&
$ 9.0434 (-2) $\\
\hline
\end{tabular}
\end{table}

\subsubsection{High-$ Q $ regime}
In the high $ Q $ regime, solutions have $ Q\geq Q_{\rm th}=7.35\times
10^{-4} $. Once again, all the solutions have an isothermal,
flat-density core, except for an extremely high $ Q $ case.
In the high-$Q$ regime, however, as $ Q $ (or $ \sigma  $) increases, the
core density increases (Fig. [\ref{fig-high}]; Table \ref{solution}),
contrary to the behaviour 
observed for the low-$Q$ regime. Therefore, $ Q_{\rm th} $ gives
the solution with the minimum possible core density 
$ \rho_{\rm core}\simeq 10^{4}\rho _{b} $.
 
This behaviour occurs because the core of the halo is now in the short
mean free path regime. In this case, contrary to the low-$Q$ regime, the
hybrid conduction term, equation (\ref{flux}),
converges to the expression valid in the short mean free path
limit, equation (\ref{f-short}). An increase in $Q$ or $\sigma$,
therefore, results in a decrease in $\textbf{f}$. Physically,
the mean free path decreases as $Q$ or $\sigma$ increase, and it
results in reducing the heat conduction.

For an extremely high $ Q $ (or
$ \sigma  $), we found that the solution becomes identical to
that of the adiabatic infall case. This result agrees qualitatively
with the result by \citet{yoshida-strong} and \citet{moore-sidm},
where they performed a smoothed particle hydrodynamics simulation, 
corresponding to the fluid limit of an infinite 
cross section. What they found in haloes was a
density cusp instead of a flat-density core. Quantitatively, our result
does not fully agree with their result in which they find a profile
even steeper than that of the collisionless case. As pointed out by
\citet{yoshida-strong}, this may be attributed to the fact that
small-scale shocks arise in the case of the fluid regime, thus
increasing the entropy and ultimately steepening the central
density. Our model is based upon an assumption that mass
accretion is smooth and, therefore, cannot reproduce this effect.

\subsection{Meaning of the collisionality parameter $Q$}
\label{why_sect}
We have showed that solutions are parametrized by the
collisionality parameter $Q$, and also that there exist two regimes
divided by a threshold value $Q_{\rm th}$. To understand the physical
meaning of $Q$, we now describe two relevant quantities: mean
free path and the number of scatterings an SIDM particle experiences
per Hubble time.

\begin{figure}
\includegraphics[width=84mm]{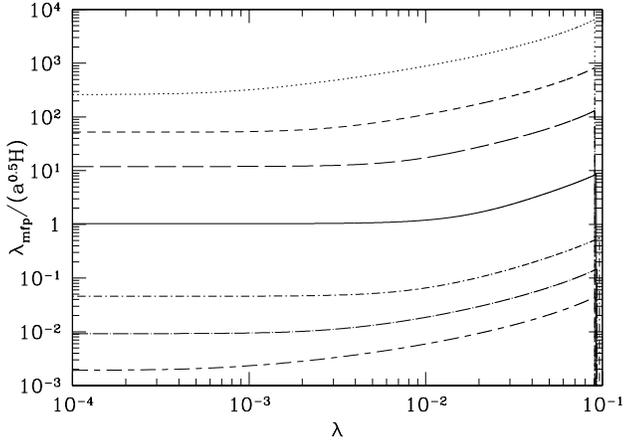}
\caption{The parameter $\eta$ versus
  dimensionless radius $\lambda$, 
  where $\eta = \lambda_{\rm mfp}/(\sqrt{a}H)$, the ratio of mean free
  path to gravitational scale height, for different
  values of $Q$. From top to
  bottom, each curve corresponds to $Q/Q_{\rm th}=1.23\cdot 10^{-3},\,
9.77 \cdot 10^{-3}, \,
6.16 \cdot 10^{-2}, \,
1, \,
17, \,
59.4$,
and 186, respectively. The constant $a=2.26$.}
\label{fig-length}
\end{figure}

We first show that the ratio of the mean free path to the gravitational
scale height (to be explained below) in the centre is very closely
related to $Q$, and it 
provides a very clean explanation of the behaviour of the similarity solution.
In previous sections, we explained the opposite trends observed in the
two different regimes 
in terms of the heat flux $\textbf{f}$ (equ. [\ref{flux}]).
The dependence of $\textbf{f}$ is determined by the ratio 
$ \eta^2 \equiv (4\pi G/p) / (a \sigma^{2})$: 
$\textbf{f} \propto \sigma$ for $\eta^2 \gg 1$, and
$\textbf{f} \propto 1/\sigma$ for $\eta^2 \ll 1$. We find that,
indeed, $\eta$ is the ratio of two length scales, the mean free path
and the gravitational scale height. The gravitational scale height (BSI),
\begin{equation}
\label{sh}
H\equiv \sqrt{\sigma_{\rm V}^{2}/(4\pi G \rho)} = \sqrt{p/(4\pi G \rho^{2})},
\end{equation}
is a rough measure of the size of a given self-gravitating system,
where $\sigma_{\rm V}^2$ is the velocity dispersion 
-- also note, however, that equation~(\ref{sh}) determines the {\it local}
gravitational scale height.
In terms of the mean free path $\lambda_{\rm mfp}\equiv 1/(\rho \sigma)$
and $a=2.26$ as in equation (\ref{f-long}),
$\eta=\lambda_{\rm mfp}/(\sqrt{a}H)$. 
We find that $\eta=1$ at the centre for $Q=Q_{\rm th}$. $\eta$ at
the centre
monotonically increases as $Q$ increases. The low-$Q$ regime then
corresponds to a condition $\eta > 1$, and the high-$Q$ to a
condition $\eta < 1$. Dependence of $\eta$ on $Q$, as well as
its radial variance, is plotted in Figure \ref{fig-length}.

\begin{figure}
\includegraphics[width=84mm]{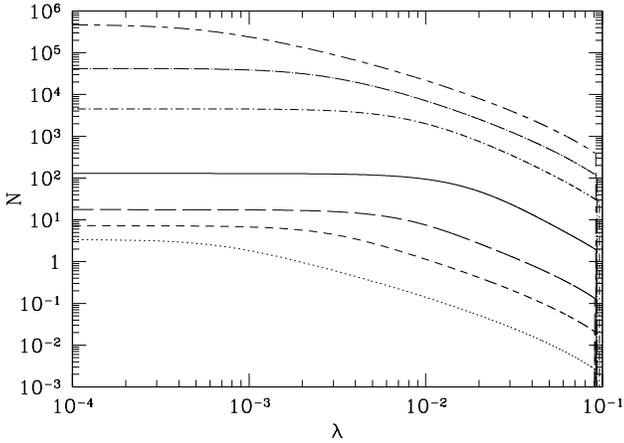}
\caption{Number of scattering that an SIDM particle
  experiences during the age of the universe in dimensionless radius
  $\lambda$. From bottom to top, each 
  curve corresponds to $Q/Q_{\rm th}=1.23\cdot 10^{-3},\,
9.77 \cdot 10^{-3}, \,
6.16 \cdot 10^{-2}, \,
1, \,
17, \,
59.4$,
and 186, respectively.}
\label{fig-qn}
\end{figure}

The number of scatterings that an SIDM particle experiences during the
age of the universe, which we denote by $N$, is also an interesting
quantity. As higher $Q$ means more frequent scattering, $N$ also
shows a monotonic dependence on $Q$ (Fig. [\ref{fig-qn}]) as $\eta$ does.
According to equation (\ref{qn}),
it can be shown that $N$ is expressible in terms of dimensionless
quantities as
\begin{equation}
\label{qn1}
N=N_{0, \rm th} \sqrt{ \frac{DP}{(DP)_{0,\rm th}}}
\left(\frac{Q}{Q_{\rm th}} \right)
 =a\sqrt{DP}\,Q/\lambda_{\rm s},
\end{equation}
where the subscript ``0,th'' refers to the value at $r=0$ for
$Q=Q_{\rm th}$, and $N_{0, \rm th}=129$.
It is interesting to see that $N_{0, \rm th}\approx 100$ is required to
achieve the maximal conductivity, namely $Q=Q_{\rm th}$. We will handle
the significance of this quantity in \S \ref{nbody_sidm_sect}, when we
compare our result to SIDM N-body simulation results. 

Finally, we stress the importance of the radial variance of
$\eta$ and $N$. When $Q/Q_{\rm th} \ll 1$ or $Q/Q_{\rm th} \gg 1$,
the system can be said, in a global sense, to reside in the long mean
free path limit or in the short mean free path limit,
respectively. When $Q/Q_{\rm th} \approx 1$, however, such a global
definition is not valid. In the $Q=Q_{\rm th}$ solution, for instance,
$\eta=1$ at the centre while $\eta \approx 10$ at the shock. 
$N$ varies from $\sim 100$ at the centre to $\sim 2$ at
the shock. This example shows that a global assumption of the short (or
long) mean free path limit is not always valid, which requires a more
careful attention when $Q\simeq Q_{\rm th}$. We will handle this
issue again in \S \ref{discussion_sect}, with respect to the estimate
on the ram-pressure stripping of substructure in a cluster environment
made by \citet{fl} and \citet{natarajan}.

\subsection{Importance of cosmological infall}
The SIDM core grows in size as a fixed fraction of the turnaround radius,
as guaranteed from the beginning because of the self-similarity of the
system.

Therefore, this model shows that cosmological infall at a
certain rate can 
completely inhibit the gravothermal catastrophe of the core by
constantly pumping hot material into the halo (see Fig. [\ref{fig-simb}]). 
This contradicts the prediction made in previous
studies
that the core would suffer gravothermal catastrophe in a Hubble
time (e.g. \citealt{bur2}). 
Even when the infall rate is smaller than that required for our
similarity solution ($ M \propto t^4 $), it will inhibit the core
collapse to some extent.
Only when the infall rate drops to a point where a system can be 
considered isolated will previous estimates of the timescale of collapse 
be valid. The net effect is a substantial delay of the
collapse phase.

\subsection{Application}
\subsubsection{Collisionality parameter as a function of $\sigma$
and $M$}
\label{qsm}
So far, we have described solutions in terms of $Q$. Now we seek a
way to apply our solutions to practical problems. This is done by
obtaining the dependence of $Q$ on the
scattering cross section $\sigma$ and the mass of haloes $M$. From
the Press-Schechter formalism, we find $Q=Q(\sigma, M)$ for
``typical'' haloes of mass $M$, which collapse
when $\sigma_{M}=\delta_{\rm crit}$ ($\sigma_{M}$ is the standard
deviation of the density fluctuations at the collapse epoch
$z_{\rm coll}(M)$ according to linear perturbation theory after the
density field is filtered on the scale $M$; $\delta_{\rm crit}$ is the
value of overdensity linearly extrapolated to a moment when the nonlinear
overdensity becomes infinite.)
Usually these are
called 1-$\sigma_{M}$ fluctuations.

As the $\varepsilon=1/6$ adiabatic infall solution has a shock at
$r_{564}$, the mass contained inside $r_{564}$, $M_{564}$, is
smaller than $M_{200}$, which is typically quoted in the
literature. In other words, the shock location is displaced
from $r_{200}$ substantially. As $r_s$ is a function of $M$ and
$z_{\rm coll}$, in order to get $Q=Q(M, z_{\rm coll})$, we first should
relate $M_{564}$ to $M$. 
We therefore need a model whose
density profile extends at least to $r_{200}$.
We use the ``truncated
isothermal sphere'' (TIS) model (\citealt{tis1}; \citealt{tis2}) for
this purpose, for the following reasons: (1) it has a unique
density profile, (2) all physical quantities are fully determined by
the values of
$M$ and $z_{\rm coll}$, (3) it has been proven to agree well with
CDM prediction in many aspects (\citealt{tis3}), and (4) the average quantity
(i.e. the average temperature) inside its own $r_{564}$ is in
a good agreement with that of our similarity solutions and of CDM
N-body haloes\footnote{We find the same level of agreement between our
similarity solutions and the N-body haloes as was found between the
TIS and N-body haloes in \citet[][see the mass-temperature relation in
  \S~8.4]{tis1}.}. 
We use a convenient set of formulae
given by \citet{tis2} to get 
$r_{\rm s}=r_{564}(M)$ for a given halo of mass $M$ at its collapse
epoch. In this model, the mass of the halo $M$ is enclosed by the
``truncation radius'' $r_t$, given by
\begin{equation}
\label{rt}
r_{t}=187.2 \left( \frac{M}{10^{12}h^{-1}M_{\sun}}\right)^{1/3}
\Omega_{0}^{-1/3} (1+z_{\rm coll})^{-1} h^{-1} \rm{kpc},
\end{equation}
and we find that
\begin{equation}
\label{m564}
M_{564}=0.587 M,
\end{equation}
and
\begin{eqnarray*}
r_{564}= 0.514 \, r_{t}
\end{eqnarray*}
\begin{equation}
\label{r564}
 = 96.22 \left(
\frac{M}{10^{12}h^{-1}M_{\sun}}\right)^{1/3}
\Omega_{0}^{-1/3} (1+z_{\rm coll})^{-1} h^{-1} \rm{kpc}.
\end{equation}
Note that $M$ in this model is equivalent to $M_{130}$.
Equations (\ref{rt}), (\ref{m564}), and (\ref{r564}) are valid only when the
universe is in the matter-dominated era. The mean matter density is in
general given by
\begin{equation}
\label{rhob}
\rho_{b}(z)=\Omega_{0}\rho_{\rm 0,crit}(1+z)^3
=1.88\times10^{-29} \Omega_{0} (1+z)^3 h^2 {\rm g /cm^{3}}.
\end{equation}

\begin{figure}
\includegraphics[width=84mm]{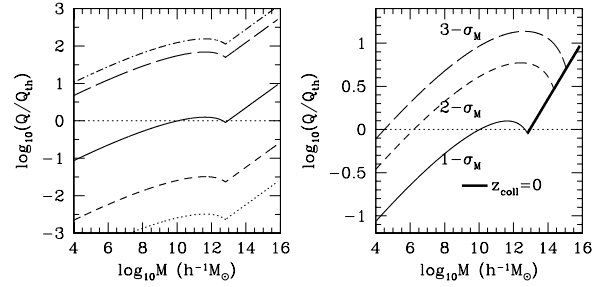}
\caption{Left: $Q$ vs. mass of haloes at their typical formation epoch for
different $\sigma $. From bottom to top, curves correspond to 
$\sigma=0.56,\, 5.6,\, 218.5,\, 1.2\times 10^{4},\, 2.7 \times 10^{4}$
respectively. They all correspond to 1-$\sigma_{M}$ density peaks
($\sigma_{M}$ means the standard deviation of the density
fluctuations filtered on mass scale $M$ at the collapse epoch); 
Right: $Q$ vs. mass of haloes at
their formation epoch for $\nu$-$\sigma_{M}$ ($\nu$=1, 2, 3)
fluctuations 
with $\sigma = 218.5 \, \textrm{cm}^{2} \textrm{g}^{-1} $. Cluster-sized haloes observed at
present will be clustered around the crossing point of the 3-$\sigma_{M}$
line and $z_{\rm coll}=0$ line. 
}
\label{fig-q}
\end{figure}

When the scattering cross section
$\sigma$ is a constant, $Q$ also remains constant in a
matter-dominated era where $\rho_{b}\propto t^{-2}$ if $\varepsilon=1/6$
(or $r_{\rm s}\propto t^{2}$). 
From the equations above, we find that
\begin{eqnarray}
\label{q-general}
Q &=& Q_{\rm th} \left(\frac{\Omega_{0}}{0.27}\right)^{2/3}
\left(\frac{\sigma}{218.5\, \textrm{cm}^2 \textrm{g}^{-1}}\right)
\left(\frac{M}{10^{10}h^{-1}M_{\sun}}\right)^{1/3} \nonumber \\
& &\times \left(\frac{h}{0.7}\right)
\left(\frac{1+z_{\rm coll}}{1+2.09}\right)^2.
\end{eqnarray}
Note that typical
(1-$\sigma_{M}$ fluctuation) haloes of
$M=10^{10}h^{-1}M_{\sun}$ collapse at $z_{\rm coll}=2.09$ in the
currently-favoured $\Lambda$CDM universe with $h=0.7$,
$\Omega_{0}=0.27$, $\Omega_{\Lambda}=0.73$ and $\sigma_{8}=0.9$. 
If we restrict the mass range of haloes such that the high mass end
will still collapse in the matter-dominated era -- $z_{\rm coll}\ga 1$
-- and the low mass end roughly satisfies the self-similarity, we
should choose haloes with masses $M\simeq[10^{6}-10^{12}]\,h^{-1}M_{\sun} $.
In this mass range, equation (\ref{q-general}) reads
\begin{equation}
\label{q-lcdm}
Q\simeq [1.68-9.31]\times 10^{-4}
\left( \frac{\sigma}{218.5\, \textrm{cm}^{2}\textrm{g}^{-1}} \right)
\end{equation}
for a $\Lambda$CDM universe
(see Fig. [\ref{fig-q}]: for $\sigma=218.5\, \textrm{cm}^{2}\textrm{g}^{-1}$, $Q=1.68\times
10^{-4}$ corresponds to $M=10^{6}h^{-1}M_{\sun}$, while $Q=9.31\times
10^{-4}$ corresponds to $M=10^{11.62}h^{-1}M_{\sun}$). Here we applied
the Press-Schechter formalism to obtain $z_{\rm coll}(M)$.
However, just for comparison, we also calculated $Q$ for haloes with
$M \ga 10^{12}h^{-1}M_{\sun}$, which typically have collapsed only
recently. 
Such massive haloes must be rare, high-$\sigma$ fluctuations in order
to collapse by the present, and
we cannot apply our similarity model to these haloes because their
mass assembly deviates substantially from self-similar evolution. Note
that the
rarer the objects are, the higher the collapse redshift $z_{\rm coll}$
is, which in turn makes $Q$ larger in equation (\ref{q-general}). This trend
can be seen in Figure \ref{fig-q}.

\subsubsection{Comparison with N-body simulation results}
\label{nbody_sidm_sect}
Consider the range of $\sigma$ identified in
the cosmological N-body simulation of SIDM by
\citet{dave}: $ \sigma = [0.56-5.6]\,\textrm{cm}^{2}\textrm{g}^{-1} $.
In a $\Lambda$CDM universe, this range of $\sigma$ in equation
(\ref{q-lcdm}) then yields the range
%% labmda_s=0.090434 for Q=6.2e-7
%% lambda_s=0.0905-0.0906 for Q=3.6e-5
\begin{eqnarray}
Q&\simeq&[4.31\times 10^{-7}-2.39\times 10^{-5}] \nonumber \\ 
&\simeq&[5.86\times 10^{-4}-3.25\times 10^{-2}]\,Q_{\rm th}
\label{q-accept}
\end{eqnarray}
for haloes with $10^{6}h^{-1} M_{\sun} <M<10^{12} h^{-1}
M_{\sun}$. Here $Q=4.31\times 10^{-7}$ corresponds to $\sigma=0.56
\, \textrm{cm}^{2}\textrm{g}^{-1}$ and $M=10^{6}h^{-1}M_{\sun}$, while
$Q=2.39\times 10^{-5}$ corresponds to $\sigma=5.6
\, \textrm{cm}^{2}\textrm{g}^{-1}$ and $M=10^{11.62}h^{-1}M_{\sun}$.

The simulation results of \citet{dave} reside in the low-$Q$ regime, as
seen in equation (\ref{q-accept}). Our solutions allow us to identify
a corresponding range of $Q$-values in the high-$Q$ regime for
which the density profiles are indistinguishable from their low-$Q$
counterparts.
The range of solutions described by equation
(\ref{q-accept}) can be matched by solutions with
\begin{equation}
\label{q-accept-high}
Q\simeq [1.37\times 10^{-2}-0.17]\simeq [18.6-231]\,Q_{\rm th},
\end{equation}
or $ \sigma \simeq [1.2\times 10^{4}-2.7\times 10^{4}]\,
\textrm{cm}^{2}\textrm{g}^{-1} $. 
Here $Q=1.37\times 10^{-2}$ corresponds to $\sigma=1.2\times
10^{4} \, \textrm{cm}^{2}\textrm{g}^{-1}$ and $M=10^{6}h^{-1}M_{\sun}$, while
$Q=2.7\times 10^{-2}$ corresponds to $\sigma=2.7\times 
10^{4} \, \textrm{cm}^{2}\textrm{g}^{-1}$ and $M=10^{12}h^{-1}M_{\sun}$.

Therefore, we predict that a cosmological N-body simulation of SIDM 
with $\sigma \simeq [1.2\times 10^{4}-2.7\times
  10^{4}]\, \textrm{cm}^{2}\textrm{g}^{-1} $ will 
produce results similar to those of the \citet{dave} simulations.
The nondimensional core density $ D(0) $ corresponding to
the range of $Q$ found in this section
lies in the range $D(0) \simeq [3.3\times 10^{4}-7.4 \times 10^{5}]
$, while the nondimensional 
core temperature $ T_{\rm core}\equiv \frac{3P(0)}{2D(0)} $ is in the
range $ T_{\rm core}\simeq
[0.11-0.41] $. However, in order to obtain observationally acceptable
values, we should actually fit the implied rotation curves of our similarity
solutions directly to the empirical data. This is the main topic of
\S \ref{rotation_sect}. 

\subsubsection{Rotation curve fitting}
\label{rotation_sect}
In this section, we find best-fitting similarity solutions
which match the observed rotation curves of dwarfs and LSBs. To do so, we
simply compare our similarity solutions to an empirical fit by
\citet{bur1}. Burkert found that a density profile given by
\begin{equation}
\label{burkert-density}
\rho(r)=\frac{\rho_{0} r_{0}^{3}}{(r+r_{0})(r^2+r_{0}^{2})},
\end{equation}
where $ \rho_0 $ and $ r_0 $ are free parameters which represent
the central density and a scale radius, respectively, matches halo
density profiles which are derived from the observed rotation curves
of dwarf galaxies. Recent studies of high-resolution H$\alpha$ rotation curves
of dwarfs and LSBs confirm this result (\citealt{march}; see also
references therein): they use a ``hybrid''  of
H$\alpha$ and HI rotation curves which can extend from the centre to
$r_{\rm{max}}$, and find that the Burkert profile is the best fit
for this range.

\begin{figure*}
\includegraphics[width=160mm]{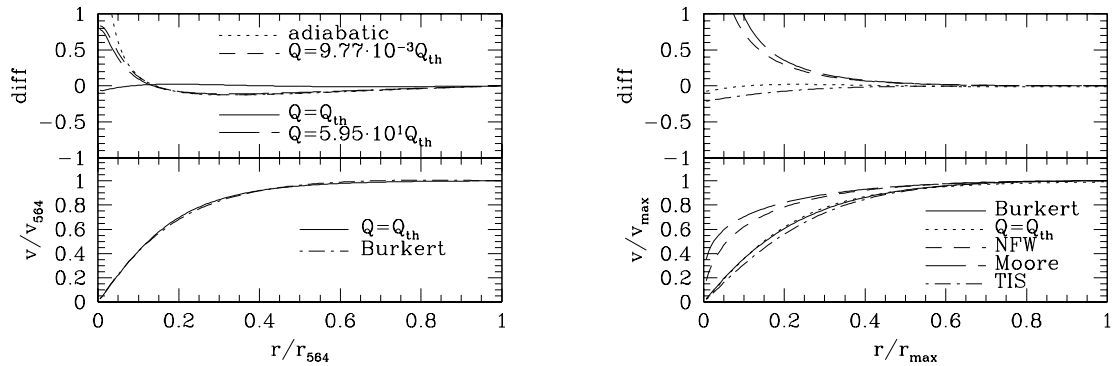}
\caption{Rotation curve fitting. The upper left panel compares several
  different $Q$-solutions to the Burkert profile, normalized to
  $r_{564}$. The lower left panel
  shows the best-fitting solution to the Burkert profile, namely the
  $Q_{\rm th}$ solution. 
  The right panel compares various halo models to
  the Burkert profile, normalized to $r_{\rm max}$.
  It has the $Q_{\rm th} $ profile in $r$ and $v$ in units of $r_{\rm
  max, Burkert}= 0.835 \, r_{564, {\rm SIDM}} $ and
  $ v_{\rm max, Burkert} = 1.01 \, v_{564, {\rm SIDM}} $, respectively,
  for the same profile as plotted in the lower left panel.
  In both boxes, the top panel shows the relative difference of a
  given profile, $(v-v_{\rm Burkert})/v_{\rm Burkert}$. The line types
  of the upper right panel follows the meaning of those in the lower
  right panel.}
\label{fig-rotation}
\end{figure*}

For fitting purposes, one should specify a radius where the circular
velocity ($v=\sqrt{Gm/r}$) is the same for different halo
models. In other words, local 
density may vary but the
mass enclosed by such a 
radius -- let us denote it by the ``normalization radius'' $r_{n}$
-- should be the same. We choose
$r_{n} = r_{564}$, the ``shock radius'' of our similarity solutions,
to find the best-fitting 
similarity solution to the Burkert 
profile. 
In this case, the ``concentration parameter''
$c_{564, {\rm Burkert}}\equiv r_{564}/r_{0}$ is the only free
parameter. The goodness of a fit is observed through the relative mean
square deviation, 
\begin{equation}
\chi^{2}/\nu \equiv \sum_{i} \left( \frac{v(r_{i})-v_{\rm
    Burkert}(r_{i})}{v_{\rm Burkert(r_{i})}} \right)^{2}/N, 
\end{equation}
where $r_{i}$ is the radius of the $i$th data point and $N$ is the number
  of such points.

\begin{table}
\caption{Best-fitting concentration parameter of the Burkert profile and
  $\chi^2/\nu$. $\chi^2/\nu$ is normalized by the value 
  found for $Q_{\rm th}$ solution, $(\chi^2/\nu)_{Q_{\rm th}}=2.06\times
  10^{-4}$, the minimum.
}
\label{rotationfitting}
\begin{tabular}{ccccc}
\hline 
$ Q/Q_{\rm th} $&
$c_{\rm 564, Burkert}$ &
$(\chi^2/\nu)/(\chi^2/\nu)_{Q_{\rm th}}$ \\
\hline
$ 0 $&
$ 5.51$&
$ 4.72 (2) $\\
$ 1.23 (-3) $&
$ 5.51$&
$ 3.86 (2) $\\
$ 9.77 (-3) $&
$ 5.49$&
$ 1.36 (2) $\\
$ 6.16 (-2)  $&
$ 4.98$&
$ 2.51 (1) $\\
$ 1 $&
$ 3.95$&
$ 1 $\\
$ 1.70 (1) $&
$ 5.17$&
$ 1.11 (1) $\\
$ 5.95 (1) $&
$ 5.64$&
$ 9.82 (1) $\\
$ 1.86 (2) $&
$ 5.59$&
$ 2.99 (2) $\\
$ \infty  $&
$ 5.51$&
$ 4.72 (2) $\\
\hline
\end{tabular}
\end{table}

We find that the solution with $Q=Q_{\rm th}$ is best fit to the Burkert
profile, as seen clearly in Figure \ref{fig-rotation} and Table
\ref{rotationfitting}\footnote{We find the same answer, that the
  $Q_{\rm th}$ solution is the best-fit 
  to the Burkert profile, when we relax the constraint
  $r_{n}=r_{564}$ and assume that the two parameters,
  $\rho_{0} $ and $r_{0}$, are both free.}\footnote{The
  values of $\chi^{2}/\nu$ shown in 
  Table~\ref{rotationfitting} show the relative quality of the fits of
  the Burkert profile to our self-similar SIDM profiles for different
  values of $Q$. In order to interpret these values in an absolute
  sense to determine if the fits are acceptable, we need to know the
  uncertainty of the fit of the Burkert profile to the observational
  data points. For example, if the 1-$\sigma$ error bars are all a
  fraction $f$ of the value of the data points, then the quantity
  $f^{-2} (\chi^{2}/\nu)$ should be small, i.e. $\la 1$, for the
  theoretical curves to be a good fit at the 1-$\sigma$
  level. Since $(\chi^2/\nu)_{Q_{\rm th}}=2.06\times
  10^{-4}$, the SIDM profile
  for $Q=Q_{\rm th}$ is a good fit to the data as long as $f\ga 0.014$.
}.
As shown in the right panel of Figure~\ref{fig-rotation}, when
$Q=Q_{\rm th}$, the SIDM halo rotation curve is virtually
indistinguishable from the Burkert profile at all radii $r\le r_{\rm
  max}$, by contrast with the strong disagreement at small radii
between the Burkert profile and the NFW and Moore profiles, respectively.
Since the $Q=Q_{\rm th}$ solution has the most effective conductivity,
which is closest to an isothermal structure among the similarity
solutions, we argue that dwarfs and LSBs described by the Burkert
profile are systems which are almost fully relaxed. This argument is
supported by the fact that the TIS solution is almost
identical to the Burkert profile (see Fig. [\ref{fig-rotation}]; also
refer to \citealt{tis2}). 
The TIS solution is obtained by assuming
that the system has a uniform temperature, isotropic random velocity,
and the minimum possible energy, which in effect is equivalent to
assuming a fully relaxed system. The solution with $Q=Q_{\rm th}$
corresponds to 
the most relaxed system among our similarity solutions.

\subsubsection{High value of $\sigma$: Contradiction with SIDM N-body
  simulation results?}
An attempt to identify the range of SIDM cross section required to
produce density profiles in agreement with dwarf galaxy rotation
curves was previously made using N-body simulations of SIDM halo
formation from Gaussian-random-noise cosmological density fluctuations
(\citealt{dave}; \citealt{yoshida-weak}). These N-body results also
indicated that, for this range of SIDM cross section, larger mass
haloes (i.e. from the Milky Way to clusters) produce density profiles
with flattened cores which are even more pronounced than those of
dwarf galaxies. Since astronomical evidence suggests that such large
haloes have relatively smaller cores, if any, than dwarf galaxies, this
has led to the suggestion that the SIDM cross section must depend upon
the relative velocity of the scattering events, decreasing with
increasing velocity to suppress this effect in larger haloes
(e.g. \citealt{cavf}). How do our self-similar solutions for SIDM halo
formation compare with these N-body results?

The value of $Q$ in the similarity solutions which best fits the dwarf
galaxy rotation curves, 
$Q_{\rm th}$, corresponds to a value of the SIDM cross section when we
identify the halo mass and redshift to which we apply the similarity
solution, as described above in \S~\ref{qsm}. 
For haloes of mass $M\simeq
10^{10} h^{-1} M_{\sun}$, which roughly represents the mass of dwarfs
and LSBs observed, the solution with $Q=Q_{\rm th}$ implies that
$\sigma\simeq 218.5 \, \textrm{cm}^{2} \textrm{g}^{-1}$, if the
observed galaxies formed at the typical epoch for their mass scale
(i.e. 1-$\sigma_M$ fluctuations).
This value is significantly larger than the range of acceptable cross
section values reported for the N-body results for SIDM haloes by
\citet{dave} (equ. [\ref{q-accept}]) and similar results by
\citet{yoshida-weak}. 

Such discrepancy is observed also in $N$, the number of scatterings
that an SIDM particle experiences during the age of the
universe. \citet{yoshida-weak} report that $N\simeq 1 - 10$ is enough
to generate a soft core in their N-body SIDM haloes.
We have seen in \S~\ref{rotation_sect}, however, that $Q\approx
Q_{\rm th}$ is required to find acceptably flattened soft cores, and this in
turn corresponds to $N\simeq 100$ in the core region as described in
\S~\ref{why_sect}. 

We may attribute this discrepancy to the fact that 
\citet{dave} did not actually perform rotation
curve fitting with their N-body results. \citet{dave} instead
found their preferred value of $\sigma$ by constraining the halo density at
$r=1$kpc for haloes at the present epoch
to be in the range $[0.01-0.1]M_{\sun}{\rm{pc}}^{-3}$. 
However, limited numerical resolution prevented them from determining
the halo density at radii as small as those required to match the
observed rotation curves which show flat-density cores. Our results
suggest that, had their simulations been capable of resolving the
profile at smaller radii, they would have found that the density
continued to rise to a higher value at smaller radii.
As seen in Figure 1 of \citet{dave}, three
haloes with $M\simeq 10^{9} M_{\sun}$ ($M=(1.7, 0.9, 1.1)10^{9}
M_{\sun}$) are almost 
unaffected by the inclusion of SIDM collisionality if
$\sigma\simeq[0.56-5.6]\,\textrm{cm}^{2}\textrm{g}^{-1}$, which
implies that the 
NFW-type cuspy profile persists in these haloes despite the SIDM interaction.

Conclusions drawn from current N-body simulations of SIDM halo
profiles may, therefore, require revision in light of our results.
The scattering cross-section $\sigma\simeq218.5 \, \textrm{cm}^{2}
\textrm{g}^{-1}$ is in an interesting regime which has not been studied
before by N-body simulations. This value may also help to resolve the
problem identified by the N-body simulations for smaller cross
section, in which larger-mass haloes result in relatively larger
cores. The small cross section regime corresponds to $Q < Q_{\rm th}$,
for which the effect of SIDM conduction increases with increasing
$Q$. As shown in Figure~\ref{fig-q}, larger-mass haloes typically have
larger $Q$-values, since their larger sizes more than offset the lower
mean densities which result from their later formation. According to
our results, however, haloes from dwarf galaxies to clusters are not
in the small cross section regime (i.e. low-$Q$ regime). In fact,
the dwarf galaxy rotation curves prefer $Q=Q_{\rm th}$, so large mass
galaxies and clusters have $Q>Q_{\rm th}$, in general (see
Fig.~[\ref{fig-q}]). According to Figure~\ref{fig-high}, this high-$Q$
regime suppresses conduction, yielding smaller cores (i.e. higher
central densities) for higher-mass haloes. We predict, therefore, that
as long as $\sigma\simeq218.5 \, \textrm{cm}^{2}\textrm{g}^{-1}$ is
used in an SIDM N-body simulation, a constant value, independent of
velocity, will suffice to match both dwarf galaxy rotation curves and
the mass profiles of larger-mass haloes.

\section{Conclusion/Discussion}
\label{discussion_sect}
CDM particles have been assumed to be collisionless in the standard
CDM theory of cosmic structure formation. Despite the success of this
standard theory, the elementary particle physics theory necessary to explain
the origin and microscopic properties of the particles which comprise
this dark matter is not yet known. It is natural for us to ask if the
microscopic nature of CDM particles might lead to further constraints
on this theory by astronomical observation. We have focused here on
one such microscopic property, that of self-interaction by elastic
scattering, and its effect on the internal structure and dynamical
evolution of virialized CDM haloes during galaxy and large-scale
structure formation.
The apparent discrepancy between the observed
density profiles of the haloes of dark-matter dominated dwarf and LSBs and those
predicted by N-body simulations of collisionless CDM may well be
resolved if one assumes that CDM particles interact with each other
non-gravitationally. The self-interacting dark matter (SIDM)
hypothesis is an attempt to produce such a nongravitational
interaction between dark matter particles.

We have derived the first fully-cosmological similarity solutions for
halo formation in the presence of collisionality. This provides an
analytical theory of the effect of the self-interacting dark matter
(SIDM) hypothesis on CDM halo density profiles as follows:

\begin{itemize}
\item We have adopted the spherical infall model of cosmological halo
  formation, guided by the results of N-body simulations of CDM. The
  collisional Boltzmann equation for the evolution of the gas of CDM
  particles yields a set of fluid-like conservation equations under
  the assumptions of spherical symmetry and isotropic velocity
  distribution. The effect of self-interaction collisions is accounted
  for by an effective conductivity term in the energy equation. This
  conductivity is valid for arbitrarily large or small collision mean
  free path $\lambda_{\rm mfp}$.
\item For an Einstein-de Sitter universe (or a flat universe with
  cosmological constant, at early times when matter dominates), the
  nonlinear growth of perturbations which leads to halo formation in
  the spherical infall model can be described by similarity solutions
  in the absence of conductivity. In the presence of SIDM
  conductivity, self-similarity is still possible, but only for mass
  perturbations $\delta M /M \propto M^{-1/6}$. Remarkably, this
  self-similarity required in our solution is well-motivated and
  justified by the theory of halo formation from peaks of the Gaussian
  random noise density fluctuations (\citealt{hs}). For galactic
  haloes, which form from density fluctuations
whose power spectrum can be approximated by a power law 
$P(k)\propto k^{-2.5}$, the conditions required for this particular
  self-similarity naturally arise.  
\item According to our similarity solutions, collisions of SIDM transport
heat from the hotter, outer halo region into the colder core
region. This process flattens the central density, and continuous infall
pumps energy into the halo which stabilizes the core against
gravothermal catastrophe.
\item These solutions are characterized by a single dimensionless
  quantity, the collisionality parameter $Q\equiv \sigma \rho_{b}
  r_{\rm vir} \propto r_{\rm vir}/\lambda_{\rm mfp}$, where $\sigma$ is the
  scattering cross 
section per unit mass, $\rho_{b}$ is the mean matter density, 
$r_{\rm vir}$ is halo virial radius and $\lambda_{\rm mfp}$ is the
  collision mean free path. The maximum flattening of central density
  occurs for an intermediate value of $Q$, $Q_{\rm th}=7.35\times 
10^{-4}$, at which the 
  halo is maximally relaxed to isothermality. The density profile of
  the $Q_{\rm th}$ solution
  matches that inferred from the observed rotation curves of dwarfs
  and low surface brightness 
  galaxies (LSB) very well.
\item In the low-$Q$ regime ($Q<Q_{\rm th}$), flattening of the
  central density profile becomes {\it stronger} as $Q$ increases. Previous
  cosmological SIDM N-body simulations with $\sigma\simeq [0.1 - 10]
  {\rm cm}^{2} {\rm g}^{-1}$ lie in this regime
  (\citealt{yoshida-weak}, \citealt{dave}, \citealt{cavf}).
  Central density profiles became flatter as they increased the
  scattering cross-section, which is equivalent to increase in $Q$,
  because $Q \propto \sigma $. On the
  contrary, in the high-$Q$ regime ($Q>Q_{\rm th}$), flattening of the
  central density becomes {\it weaker} as $Q$
  increases. This happens because the scattering mean free path becomes
  shorter as $Q$ increases. SIDM simulations which adopt a fully
  collisional limit 
  to derive the maximal density flattening,
  which corresponds to ordinary gas dynamics
  (\citealt{yoshida-strong}; \citealt{moore-sidm}), report that haloes
  obtain density profiles with central cusps as steep as or steeper
  than those in
  collisionless N-body simulations. This seemingly puzzling behaviour
  is easily explained: $\sigma \rightarrow \infty$ corresponds to $Q
  \rightarrow \infty$, and therefore density flattening becomes negligible.
\item Under the assumption that dwarfs and LSBs
  formed at their typical collapse epoch in $\Lambda$CDM,
  $\sigma\simeq 200 {\rm cm}^{2} {\rm g}^{-1}$ makes $Q=Q_{\rm th}$,
  much higher than previous estimates, $\sigma\simeq[0.5-5] {\rm
  cm}^{2} {\rm g}^{-1}$, based on N-body experiments. This value of
  $\sigma$, independent of halo mass, would make $Q>Q_{\rm th}$ for
  clusters, which typically formed only recently, resulting in
  relatively less flattening of their central density profile and a
  smaller core. A velocity dependent cross-section, $\sigma \propto
  1/v $, suggested by \citet{yoshida-weak} is thus unnecessary.
\item According to our similarity solutions, the solution for
  $Q=Q_{\rm th}$ represents the solution most relaxed to isothermality
  inside the virialized postshock region. It is notable, therefore,
  that the $Q=Q_{\rm th}$ solution is very similar to the nonsingular
  TIS solution of \citet{tis1} for the post-collapse equilibrium
  structure of virialized haloes. The latter yields a mass profile
  almost indistinguishable from the mass profile of the Burkert fit to
  the rotation curves of dwarf and LSB galaxies (\citealt{tis2}), as
  is the $Q=Q_{\rm th}$ SIDM profile we have derived here. This suggests
  that the TIS halo model, which assumes that haloes are isothermal, is
  a natural outcome of the dynamical formation of CDM haloes when
  conductivity causes the halo to relax maximally toward isothermality.
\end{itemize}

One may improve upon this work by performing a cosmological N-body
simulation. As our analysis is based upon self-similarity, or a constant mass
accretion rate $\frac{\partial \log M}{\partial \log a}=6$, when the
mass accretion rate deviates from this canonical rate, our analysis is
no longer valid.
Several authors have investigated a realistic halo formation history
both by an analytic approach and by N-body experiments, 
tracking
the history of the ``most massive progenitor (MMP)''. \citet{wechsler}
performed a cosmological N-body simulation and tracked the growth
of MMP mass in a $\Lambda$CDM universe.
\citet{NS} calculated the growth of MMP for a power-law power spectrum
and \citet{vandenbosch} calculated it for a CDM power spectrum, both
using the extended Press-Schechter theory. These studies show that
mass accretion starts with a fast, rapid merger and ends with smooth,
continuous accretion. This trend
is clearly seen in equation (\ref{wechsler-mass}), where the logarithmic
accretion rate is decreasing linearly with increasing scale factor
$ a $. The accretion rate obtained in this way is different from what
is expected from HS. For $ n=-2.5 $ in the matter-dominated era, we
find from equation (\ref{growth}) that
$ \frac{\partial \log M}{\partial \log t}=4 $ 
or $ \frac{\partial \log M}{\partial \log a}=6 $. 
This fast accretion rate captures only the early mass accretion epoch
given by equation (\ref{wechsler-mass}), and therefore, one
should not apply this rate to the later epoch when mass accretion
becomes negligible. This is the moment where the evolution of SIDM
haloes deviates from the self-similarity we assumed.

For cluster-mass haloes, there is another reason that our assumption
of self-similarity breaks down and should be improved in future work;
the value of $n$ which enters the self-similar infall model in
$\Lambda$CDM actually depends weakly upon halo mass. The value we have
adopted to ensure self-similarity in the presence of SIDM
collisionality, $n=-2.5$, is appropriate for the entire range of
galactic halo masses. As Figure \ref{fig-neff} shows, however, $n$
increases with mass, and for clusters with mass above $10^{14}
M_{\sun}$, $n> -2$. If $n \neq -2.5$, self-similarity is violated by
the presence of SIDM terms in the equations. Since the accretion rate
is lower if $n$ is higher (i.e. $ \frac{\partial \log M}{\partial \log
  a}= \frac{3}{n+3} $), the flattening effect of SIDM on the halo
central density profile may be lower on cluster scales than expected
from our self-similar solutions for $n=-2.5$.

\begin{figure}
\includegraphics[width=84mm]{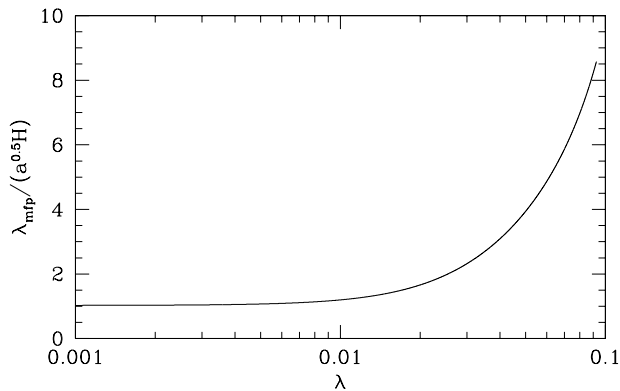}
\caption{Comparison of length scales for the $Q=Q_{\rm th}$ case. The
mean free path $\lambda_{\rm mfp}$ is comparable to the
gravitational scale height H
in the core region. However, the ratio of the mean free path to
the gravitational scale height becomes larger as the radius increases.
See also Figure \ref{fig-length}, which includes other values of $Q$.
}
\label{fig-lengthQth}
\end{figure}

Does our prediction of high value of $\sigma$, $\sigma \simeq 200
{\rm cm}^{2} {\rm g}^{-1} $, affect the abundance of dark matter
substructure? Gravitational lensing flux anomalies have been
interpreted as  evidence for the existence of dark matter
substructures in the parent halo (\citealt{mz}; \citealt{dk};
\citealt{kgp}; \citealt{mjow}). 
The self-interacting dark matter, if real, would suppress the number
of dark matter substructures to some extent. We are not sure
at this stage, however, how strong this effect will be: the simple
assumption of cold, continuous mass infall prevents us from making any
strong prediction. We instead
describe, in the following discussion, the complexity relating to
two main mechanisms for suppressing the substructure formation when dark
matter is collisional: ram-pressure stripping and thermal
evaporation. One should also note that the interpretation of the
lensing flux anomalies is not settled yet. Recent analysis indicates
that this anomaly may not require any substructure in the primary
lens halo, but only the substructure in the intergalactic space
(\citealt{metcalf}).

We now point out some caveats found in previous analyses, which also
require more realistic, cosmological analysis. As many previous
analytical estimates have been based on either isolated haloes or
very simplified models, we believe that fully cosmological N-body
simulations in the high-$Q$ ($\sigma$) regime should be carried
out to
clarify this issue. For instance, the restriction coming from the
susceptibility of SIDM haloes 
to ram-pressure stripping (\citealt{fl}; \citealt{natarajan}) may
be relaxed if we remove the simplicity in their analysis. They determine
the truncation radius of a galaxy in a cluster by the condition
\begin{equation}
\label{ram}
\rho_{c}(r)v_{g}^{2}=\rho_{g}(r_{t})\sigma_{g}^{2},
\end{equation}
where $v_{g}$ is the velocity of the galaxy inside a cluster,
$\rho_{c}(r)$ is the density of the cluster at radius $r$,
$\rho_{g}(r_{t})$ is the density of the galaxy at its truncation
radius $r_{t}$, and $\sigma_{g}$ is the internal velocity
dispersion of the galaxy. Equation (\ref{ram}) is valid for highly
collisional fluid. They then use the restriction that
\begin{equation}
\label{eq-natarajan}
\frac{\lambda_{\rm mfp}}{r_{t}}\simeq\frac{1}{\sigma \Sigma(r_{t})} \leq 1
\end{equation}
where $\lambda_{\rm mfp}$ is the collision mean free path and
$\Sigma(r_{t})$, which is $0.024 \, \textrm{g}\,\textrm{cm}^{-2}$ for
their fiducial case,  is 
the surface density of a galaxy at $r_{t}$.
However, this is a crude way of describing the fluid regime where the
equation (\ref{ram}) can be applied.
There exists a regime where the SIDM can be
treated as collisional only at the centre because 
the ratio of the mean free path to the gravitational scale
height becomes larger as the radius of a galaxy increases (see Fig.
[\ref{fig-lengthQth}]; 
when we express these length scales with dimensionless terms used for
our similarity solutions, the 
scattering mean free path is $ L_{\rm mfp} \equiv \frac{1}{DQ}$, and
the gravitational scale height is $ H \equiv \frac{\sqrt{3P/2}}{D}$).
In such cases, a ``global'' application of equation
(\ref{ram}) is not valid.
Ram-pressure stripping in this case would
not be as severe as in the case of purely collisional fluid, because
cluster SIDM particles have a high probability of penetrating an SIDM
galactic halo
deeply without experiencing collision at $r \ga r_{t}$.
Therefore, to apply equation
(\ref{ram}), we have to be more conservative in defining the fluid
regime, which will relax the constraint $\sigma \la 42 {\rm cm}^{2}
{\rm g}^{-1}$ (\citealt{natarajan}) set by equation (\ref{eq-natarajan}).

\citet{go} constrained $\sigma$ from their numerical and
semi-analytical calculation of the evaporation time of elliptical
galaxies embedded in a cluster environment. According to their
analysis, such galactic haloes can evaporate within the age of the
universe if $\sigma=[0.3 - 10^{4}]\textrm{cm}^{2} \textrm{g}^{-1}$. However, their
analysis is based on a fixed cluster environment. Consider a
cosmological variant of this problem: for instance, an elliptical
galaxy may form by a recent merger at $z=1$ when the temperature of
the cluster, which only forms at $z\simeq0$, is very low. When the
evaporation time is about the age of the universe ($\sim 10^{10}
\rm{yr}$), the elliptical galaxy has a fair chance to survive
because the evaporation time is greater than the time from its
formation epoch to the present. The excluded range of $\sigma$ would
then change substantially by shifting the marginal
values, $\sigma\simeq0.3\, \textrm{cm}^{2} \textrm{g}^{-1}$ or $\sigma\simeq 10^4\, \textrm{cm}^{2}
\textrm{g}^{-1}$, which correspond to the evaporation time of the order of the
age of the universe. Moreover, their analysis is only valid for either
the very long mean free path limit or the very short mean free path
limit. They exclude the intermediate regime at $\sigma\simeq 200\, {\rm cm}^{2}
{\rm g}^{-1}$ simply by an extrapolation of these two regimes.

Because of these problems, we assert
that a more refined, fully cosmological analysis and new
cosmological N-body simulations 
with a wider range of $\sigma$ values, including
$\sigma\simeq[100-500] {\rm cm}^{2} {\rm g}^{-1}$,
should be performed. 
Even though our 
analysis here is fully cosmological, it is restricted by the fact that it
is based on a constant logarithmic mass accretion rate
(i.e. $\frac{\partial \log M}{\partial \log a} = \frac{1}{\varepsilon}
= \frac{3}{n+3}=6$ for $n=-2.5$)
that provides self-similarity.
However, a more realistic mass accretion history constructed from
merger trees in N-body simulations shows a gradual decrease of the
logarithmic mass accretion rate over time, as seen in equation
(\ref{wechsler-mass}) (e.g. \citealt{wechsler}).
When
the mass accretion rate becomes very small at late times, the underlying
halo properties 
will deviate significantly from self-similarity. Moreover, the
relatively high scattering cross-section which we 
find provides the best-fitting to dwarf galaxy rotation curves --
$\sigma \simeq 200 {\rm cm}^{2} {\rm g}^{-1}$ -- has never been
tested in cosmological SIDM N-body simulations. We will explore these
issues further in future work.

\section*{\sc Acknowledgments}
We thank M. Alvarez, S. Shapiro, H. Martel, I. Iliev, and L. Chuzhoy
for helpful discussions.
This work was supported by NASA 
Astrophysical Theory Program grants NAG5-10825, NAG5-10826,
NNG04G177G, and Texas Advanced Research Program grant 3658-0624-1999.

\section*{Appendix}
We show how we got $n_{\rm eff}$ for different mass scales. It is
slightly different from the usual way to obtain $n_{\rm eff}$ by
differentiating the rms mass fluctuation, $\sigma_{M}$.

Typically, one has
\begin{equation}
\label{neff-sigma}
n_{\rm eff}=-3 \left( 1+ \frac{d \ln \sigma^{2}_{M}}{d \ln M}\right)
\end{equation}
where
\begin{equation}
\label{sigma}
\sigma^{2}_{M} \equiv
\frac{\left\langle (m-M)^{2} \right\rangle}{M^{2}}
=\frac{1}{2 \pi^{2}} \int^{\infty}_{0} P(k) W^{2}(kR)
k^{2} dk ,
\end{equation}
where $m$ is a mass enclosed by a sphere of radius $R$ which
also defines the {\it unperturbed} mass $M$ through
\begin{equation}
\label{tophatmass}
M=\frac{4\pi}{3} R^{3} \rho_{0},
\end{equation}
where $\rho_{0}$ is the present matter density
and the average $\langle \,\, \rangle$ 
is taken over all positions of the centre of these spheres.
This ``top-hat'' filtering results in a
window function 
\begin{equation}
\label{window}
W(X)=\frac{3}{X} (\sin (X) - X \cos (X)).
\end{equation}
It is then
straightforward to calculate $n_{\rm eff}$ as a function of $M$ using
equation (\ref{neff-sigma}).

However, we are interested in the $n_{\rm eff}$ which is valid if
one considers the initial average overdensity around density
peaks, $\Delta_{0}(R)$ (HS). $\Delta_{0}(R)$ is given by
\begin{equation}
\label{delta_0}
\Delta_{0}(R) \equiv \frac{\delta
  M}{M}=\frac{\delta_{0}}{\sigma^2}\frac{1}{2\pi^2} 
\int^{\infty}_{0} P(k) W(kR) k^{2} dk,
\end{equation}
where $M$ and $W(X)$ are defined by equation (\ref{tophatmass}) and
(\ref{window}), respectively. 
From
equation (\ref{tophatmass}) and equation (\ref{delta_0}), we can see
that for a power-law power spectrum $P(k)\propto k^{n}$,
\begin{equation}
\label{delta-mass-relation}
\Delta_{0}(R) \propto R^{-(n+3)} \propto M^{-(n+3)/3}.
\end{equation}
Therefore, one can obtain $n_{\rm eff}$ as follows:
\begin{equation}
\label{neff-hs}
n_{\rm eff}=-3\left( 1+ \frac{d \ln \Delta_{0}(M)}{d \ln M}\right).
\end{equation}

The only difference in obtaining $n_{\rm eff}$ is then the power of
the window function $W(X)$ in equations (\ref{sigma}) and
(\ref{delta_0}). One can therefore expect that there would be only a
minor difference, which we confirmed (see Fig. [\ref{fig-neff}]).
We used the fitting formula for the
transfer function by 
\citet{eh}, and worked in the $\Lambda$CDM
concordance model with $\Omega_{\Lambda}=0.73$,
$\Omega_{m,0}=0.27$, $h=0.7$, $\sigma_{8}=0.9$ and untilted
Harrison-Zel'dovich primordial power spectrum.

\end{document}